\documentclass[final,3p,times,twocolumn,number,sort&compress]{elsarticle}

\usepackage{lineno,hyperref}
\usepackage{dcolumn, multirow}
\usepackage{adjustbox}
\modulolinenumbers[5]

\journal{Nuclear Instruments and Methods in Physics Research A}
\usepackage{eurosym}
\usepackage{textgreek}
\usepackage{color}
\usepackage{xcolor}
\usepackage{float}
\usepackage{graphicx}
\graphicspath{.}
\usepackage{epstopdf}
\usepackage{graphics}
\usepackage[intlimits]{amsmath}
\usepackage[utf8]{inputenc}
\usepackage{esint}
\usepackage{svg}
  \newcolumntype{d}[1]{D{.}{.}{#1}}   
  \usepackage{subcaption}

\usepackage[T1]{fontenc}

\begin{document}
\newcommand{\prc}{Phys.~Rev.~C}
\newcommand{\prl}{Phys.~Rev.~Lett.}


\title{Study of a possible silicon photomultiplier based readout of the large plastic scintillator neutron detector NeuLAND}

\author[TUDD,HZDR]{Thomas Hensel}
\author[HZDR]{David Weinberger\fnref{1}}
\fntext[1]{Deceased.}

\author[HZDR]{Daniel Bemmerer}\ead{d.bemmerer@hzdr.de} 
\author[GSI]{Konstanze Boretzky}
\author[Z,GSI,DARM]{Igor {Gašparić}}
\author[HZDR]{Daniel Stach}
\author[HZDR]{Andreas Wagner}
\author[TUDD]{Kai Zuber}

\address[TUDD]{Technische Universität Dresden, Institut für Kern- und Teilchenphysik, Zellescher Weg 19, 01062 Dresden, Germany}
\address[HZDR]{Helmholtz-Zentrum Dresden-Rossendorf (HZDR), Bautzner Landstr. 400, 01328 Dresden, Germany}
\address[GSI]{GSI Helmholtzzentrum für Schwerionenforschung, Planckstr. 1, 64291 Darmstadt, Germany}
\address[DARM]{Technische Universität Darmstadt, Fachbereich Physik, Institut für Kernphysik, 64289 Darmstadt, Germany}
\address[Z]{Ru{\dj}er Bošković Institute, Zagreb, Croatia}

\begin{frontmatter}


\begin{abstract}
The NeuLAND (New Large-Area Neutron Detector) plastic-scintillator-based time-of-flight detector for 0.1-1.6\,GeV neutrons is currently under construction at the Facility for Antiproton and Ion Research (FAIR), Darmstadt, Germany. In its final configuration, NeuLAND will consist of 3,000 2.7\,m $\times$ 5\,cm $\times$ 5\,cm big plastic scintillator bars that are read out on each end by fast timing photomultipliers. 

Here, data from a comprehensive study of an alternative light readout scheme using silicon photomultipliers (SiPM) are reported. For this purpose, a NeuLAND bar was instrumented on each end with a SiPM-based prototype of the same geometry as a 1'' photomultiplier tube, including four 6$\times$6\,mm$^2$ SiPMs, amplifiers, high voltage supply, and microcontroller. 

Tests were done out using the 35\,MeV electron beam from the superconducting Electron Linac for beams with high Brilliance and low Emittance (ELBE) with its picosecond-level time jitter in two different modes of operation, namely parasitic mode with one electron per bunch and single-user mode with 1-60 electrons per bunch. Acqiris fast digitisers were used for data acquisition. In addition, off-beam tests using cosmic rays and the NeuLAND data acquisition scheme have been carried out.

Typical time resolutions of $\sigma \leq$ 120\,ps were found for $\geq$95\% efficiency for minimum ionising particles, improving on previous work at ELBE and exceeding the NeuLAND timing goal of $\sigma < $ 150\,ps. Over a range of 10-300\,MeV deposited energy in the NeuLAND bar, the gain was found to deviate by $\leq$10\% ($\leq$20\%) from linearity for 35\,\textmu m (75\,\textmu m) SiPM pitch, respectively, satisfactory for calorimetric use of the full NeuLAND detector. The dark rate of the prototype studied was found to be lower than the expected cosmic-ray induced background in NeuLAND.
 \end{abstract}

\begin{keyword}
NeuLAND\sep SiPM\sep scintillator \sep PMT \sep saturation \sep electron beam \sep dark rate
\end{keyword}

\end{frontmatter}

\section{Introduction}
\label{sec:Intro}

The development of silicon-based photosensors, so-called silicon photomultipliers (SiPMs) \cite{Buzhan03-NIMA} as a light sensor with excellent timing properties and comparatively low operating voltages, has provided an alternative to classical photomultiplier tubes (PMTs). In the meantime a wide range of affordable types of SiPMs has become available. However, the light readout of large scintillator based neutron detectors is still dominated by the use of PMTs \cite{Nakamura16-NIMB, Wang19-NIMA}, despite ongoing developments \cite{Reinhardt16-NIMA, Wojtowicz20-Polo}. On the contrary, the use of SiPMs has become widespread in other applications \cite{Simon19-NIMA,Boehm16-JOI,DeGuio19-JOP, Strobbe17-JOI} .

One of the new-generation neutron time-of-flight detectors is the New Large-Area Neutron Detector (NeuLAND) for the R$^3$B setup at GSI/FAIR in Darmstadt, Germany \cite{Boretzky21-NIMA, Mayer21-NIMA, Douma21-NIMA}. This detector is necessary for inverse-kinematics experiments with the detection of relativistic neutrons \cite{Roeder16-PRC,Heine17-PRC} that are important for the study of astrophysical neutron capture reactions by Coulomb dissociation. 
One of the long-term goals is to independently confirm, by direct neutron detection, the existence of the recently discovered \cite{Duer22-Nature} tetraneutron state. 

Prior to the construction of NeuLAND, in a first phase studies of an approach based on iron neutron converters and multi gap resistive plate chambers had been performed \cite{Yakorev11-NIMA, Roeder12-JINST, Elekes13-NIMA}. However, for the final construction of NeuLAND, instead a granular design using fast plastic scintillator bars of 2.7\,m length and 5$\times$5\,cm$^2$ area has been adopted. This scintillator-based design is expected to display better calorimetric properties \cite{NeuLAND11-TDR}, important for the multi-neutron detection as in the tetraneutron case \cite{Duer22-Nature}. In its final, not yet reached form, NeuLAND will contain 3,000 scintillator bars (RP-408, equivalent to EJ-200 and BC-408) instrumented with altogether 6,000 fast-timing PMTs (Hamamatsu R8619 or equivalent). Groups of 50 horizontal and 50 vertical scintillator bars will form one of a total of 30 so-called NeuLAND double planes. Currently, 40\% of NeuLAND is already complete, and extensive simulations have been carried out \cite{Douma19-NIMA,Mayer21-NIMA}.

In order to improve the long-term viability of NeuLAND by offering an alternative and more economical readout option, in parallel to the ongoing construction, a study of an alternative light readout scheme by SiPMs is ongoing. In the first part of this  study, it had already been shown that SiPMs fulfil the ambitious NeuLAND timing goal of $\sigma < 150$\,ps, at close to full efficiency \cite{Reinhardt16-NIMA}. The present work reports the second part of this study. Using a new and much simplified readout prototype, again time resolution and efficiency are determined. In addition, now also the linearity of the gain over a wide range of signal amplitudes and the rate of so-called dark triggers have been investigated.  

In the final NeuLAND detector, a 0.2-1.6\,GeV neutron will scatter with a certain probability in the plastic scintillator material and give rise to a secondary proton with similar energy, in addition to possibly secondary neutrons and $\gamma$ rays. Many of the secondary proton will have similar energy as the primary neutron and therefore be close to a mimimum ionising particle. Their interactions can therefore be studied using other minimum ionising particles; in this work, 35 MeV electrons from an accelerator and cosmic-ray muons are used to this end. As a result of the scintillation process, scintillation light of 420 nm wavelength is created in the plastic. For parts of the present study, instead of scintillation light, artificial light pulses from a 422 nm wavelength pulsed laser are used. The scintillation, or laser, light is then detected either by a classical photomultiplier or, in this work mainly, by a SiPM-based prototype.

This work is organised as follows. The new SiPM-based prototype, developed as possible replacement for a 1'' PMT, is described in \autoref{sec:prototype}. It has been studied at a number of experimental setups (\autoref{sec:Experiment}), both in-beam at the ELBE 35\,MeV electron beam and off-beam with a picosecond laser and cosmic-ray muons. In \autoref{sec:DataAnalysis}, the calibration and data reduction procedures are described. The results of the in-beam measurements on time resolution, efficiency, and linearity are presented in \autoref{sec:ResultsELBE}. The off-beam results on dark rate and muon capability are given in \autoref{sec:ResultsOffline}. A discussion is given in \autoref{sec:Discussion}, and a summary and outlook in \autoref{sec:Summary}.

\bigskip
\color{blue}

\color{black}

\section{Prototype and on-board electronics} \label{sec:prototype}
\begin{figure}[tb]
    \centering
    \includegraphics[width=\columnwidth]{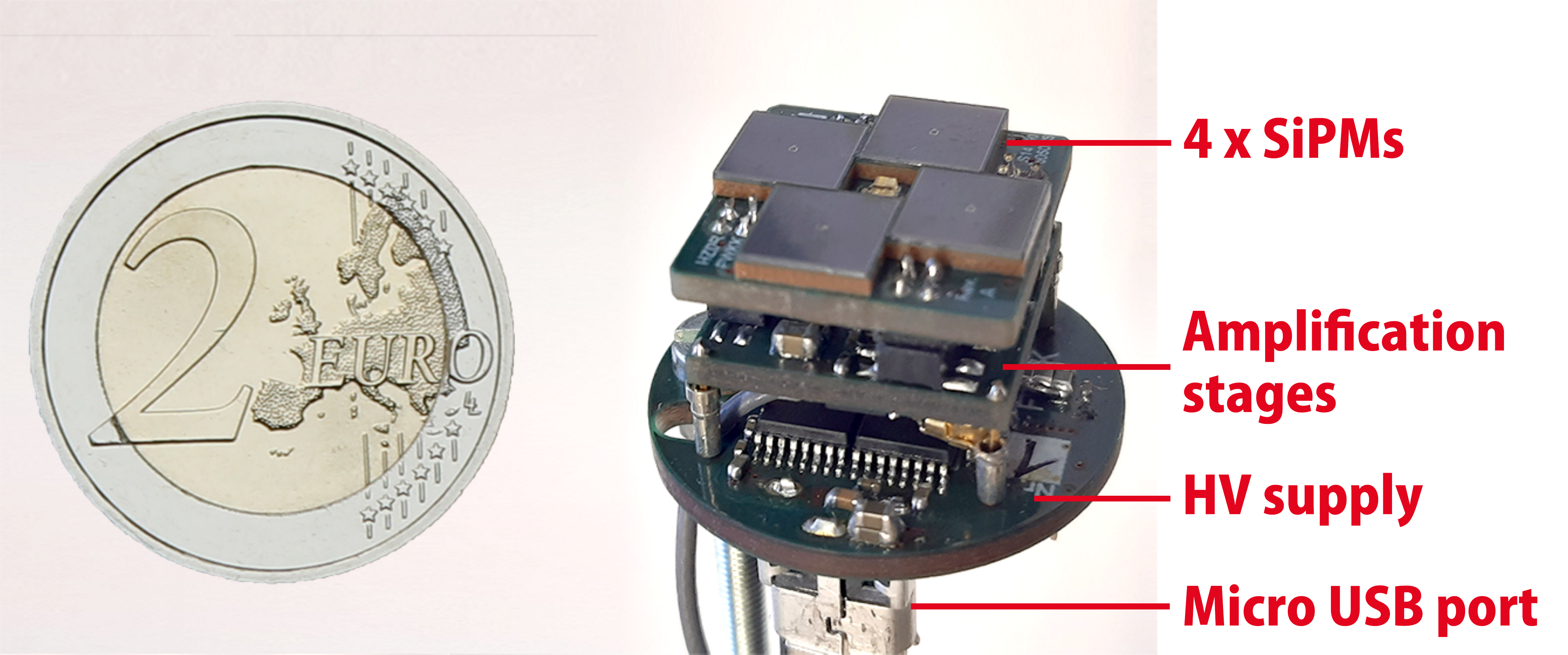}
    \caption{Photograph of the new HZDR prototype, with a coin for size comparison. The prototype fits into a 1'' diameter copper housing, to directly replace a 1'' PMT. Below the four 6$\times$6 mm$^2$ SiPMs, first the amplification stages, and, second, the high voltage supply unit are seen. The prototype is controlled via a micro USB port over that it is also powered.}
     \label{fig:prototype}
\end{figure}

Based on previous work in this series \cite{Reinhardt16-NIMA}, a new prototype was developed at  Helmholtz-Zentrum Dresden-Rossendorf (HZDR). The guiding principles were to construct a device that is easy to handle, sized to directly replace a standard 1'' PMT, and with a modular design to allow easy testing of different SiPMs types. This goal was realised with a stack of three multilayer printed circuit boards (PCBs). The stack also included a temperature-regulated SiPM bias power supply, a microcontroller, preamplification, summing, and amplification stages (\autoref{fig:prototype}). 

\begin{table*}[t!!]

\begin{tabular}{lrrlld{-1}d{-1}d{-1}d{-1}}
Type & Pitch & $n_{\rm pixels}$    & $U_{\text{BR}}$ & $U_{\text{OV}}$ & PDE & \multicolumn{1}{c}{$R_{\rm dark}$} & \multicolumn{1}{c}{Gain $g$} & \multicolumn{1}{c}{$I_{\text{dark}}$} \\
    & [\textmu m]  &    & [V]  & [V]  & [\%]  &   \multicolumn{1}{c}{[10$^6$ s$^{-1}$]} & \multicolumn{1}{c}{[10$^6$]}    & \multicolumn{1}{c}{[\textmu A]} \\ \hline

{\tt onsemi} C Series 60035 & 35             & 18980      & 24  & 2.5   & 31  & 1.2  & 3.0 & 0.7          \\
{\tt onsemi} J Series 60035 & 35             & 22292      & 24  & 2.5   & 38  & (1.9)  & 2.9 & 0.9          \\
Hamamatsu S14160-6050 & 50             & 14331      & 38    & 2.7   & 50  & (6.2)  & 2.5 & 2.5          \\
Hamamatsu S14160-6075 & 75             & 6364       & 38    & 2.7   & 57  & (7.0)  & 5.5 & 6.2          \\
Hamamatsu S13360-6075 & 75             & 6400       & 53    & 3.0   & 50  & 2.0  & 4.0 & (1.3)          \\ \hline

\end{tabular}
\caption{ $6\times6$\,mm$^2$ SiPMs used with their data sheet values. The dark rate $R_{\rm dark}$ and and dark current $I_{\rm dark}$ are converted as $R_{\rm dark} = I_{\rm dark} / (g \times e)$ with $g$ the gain and $e$ the elementary charge, when not directly given in the data sheet. Converted values are given in brackets.
}
\label{tab:sipm_types}
\end{table*}

\subsection{SiPMs studied and their arrangement in the prototype}

The prototype was instrumented with four $6\times6$\,mm$^2$ SiPMs. Its modular design allowed to study subsequently five different types of SiPMs. Their data sheet values are given in \autoref{tab:sipm_types}. 

For geometrical reasons, an array of standard square SiPMs cannot instrument the entire 1'' diameter end surface of the NeuLAND bar that initially motivated the present study. In order to study the effects caused by this limitation, a Monte Carlo simulation of the propagation of the scintillation light caused by an initial 35\,MeV electron transiting the NeuLAND bar was performed. This simulation used the GEANT4 \cite{Geant416-NIMA} package, with the {\tt QGSP\_BERT\_HP} and {\tt G4OpticalPhysics} physics lists, with scintillation and Cherenkov processes enabled  \cite{vanDerLaan10-PMB,Hartwig14-NIMA}. 
The realistic NeuLAND bar geometry was implemented, with a total length of 270\,cm, a square surface area of $5\times5$\,cm$^2$ and at each end a 10\,cm conical shaped transition from the square area to a 25\,mm diameter circular light output area. The datasheet properties of RP-408 (polyvinyltoluene, density 1.032 g/cm$^3$, refractive index $n=1.58$, 10,000 photons/MeV$_{ee}$ light yield) were adopted, and perfect reflection was assumed at the NeuLAND bar sides.  The simulation showed that the circular ends of the NeuLAND bar are homogeneously illuminated, so the precise placement of the SiPMs on the bar does not matter. 

At the center of the SiPM layout, a pulsed LED is installed, to monitor the device performance. In addition, a temperature sensor (TFPT0603L1001FV) for the SiPM bias temperature correction is placed there (\autoref{fig:prototype}). The optical coupling between the polished scintillator surface and the SiPM board was made via a foil made of optically clear silicone (QSil 219) with 1\,mm thickness.

\subsection{Base board for SiPMs, SiPM power supply, and on-board preamplification} 
\label{subsec:BaseBoard}

The device is placed on a printed circuit board (PCB) and equipped with a Cypress (now Infineon) PSoC 5LP (CY8C5888LTI-LP097) microcontroller. Via a micro-USB port the device is powered with 5\,V, and data can be exchanged with a PC. 

The SiPM high voltage is generated on the PCB by a Hamamatsu type C14156 bias power supply. The C14156 provides a bias voltage of up to 80\,V with a maximum current of 2\,mA. The temperature dependence of the SiPM breakdown voltage is compensated by the C14156 unit, based on the reading of a thermoresistor directly coupled to the SiPMs and a user-adjustable temperature compensation function. 

The output voltage and the slope of the temperature compensation are controlled by analog control voltages supplied to the C14156 by 20\,bit Texas Instruments DAC1220 Digital-to-Analog converters. The DACs, in turn, are controlled by the microcontroller.

Each of the four SiPMs is connected to its own two staged transistor (BFU550A) preamplifier. The outputs of the four preamplifiers are summed in an ADG4612 analog switch. For calibration purposes, the four ADG4612 input channels can be switched on independently by the microcontroller. The ADG4612 sum signal is finally amplified with a Texas Instruments OPA695IDBVT and routed out. An alternative mode of operation can be enabled by optionally adding an extra board between amplifier board and SiPM mounting board to directly route out the SiPM signals without any amplification.

\section{Experimental setup} 
\label{sec:Experiment}

Several different experimental setups have been used. Most experiments were done at the ELBE 35\,MeV electron beam (\autoref{subsec:ELBE}), using a dedicated electron beam setup (\autoref{subsec:Electronbeamsetup}) and digital electronics (\autoref{subsec:ExpElectronics}). In addition, off-beam calibration studies were carried out using a picosecond laser system (\autoref{subsec:ExpPiLas}), and the dark trigger rate was studied using an external trigger (\autoref{subsec:ExpDark}). Finally, the response to cosmic-ray muons was studied using the original NeuLAND data acquisition system at GSI Darmstadt (\autoref{subsec:ExpGSI}).

\subsection{ELBE direct electron beam}
\label{subsec:ELBE}

The ELBE center for high power radiation sources at HZDR\footnote{Web site \url{https://www.hzdr.de/elbe}} includes two different electron sources and a number of beam lines including direct electron beam \cite{Yakorev11-NIMA, Roeder12-JINST, Karsch12-MedPhy, Elekes13-NIMA, Kroll13-MedPhy, LASOGARCIA16-NIMA, Reinhardt16-NIMA}, bremsstrahlung \cite{Schwengner05-NIMA}, neutron time-of-flight \cite{Beyer13-NIMA}, positron, and terahertz beams. ELBE delivers up to 40\,MeV electron beam with a wide range of repetition rates and beam intensities from single electrons up to 1.6\,mA continuous wave current.

For detector tests, the direct electron beam is produced by the ELBE thermionic electron source, accelerated and deflected, and exits the evacuated beam line ($10^{-10}$\,hPa) through a thin beryllium window into air. The 13 MHz radio frequency applied to the electron gun has a jitter on the few ps level and serves as timing reference. Due to this arrangement, it is not necessary to include time reference detectors with a time resolution that is better than that of the devices under study.

There are two ELBE modes of operation, called single-user and parasitic mode, respectively. In single-user mode, the beam intensity is controlled via the operating parameters of the electronic gun, and using very low gate voltages at the extraction grid bunches of just 1-60 electrons are created and accelerated. Off-axis electrons created by discharges are absorbed in remotely controlled, movable screens along the beam line \cite{Naumann11-NIMA}. In parasitic mode, the direct electron beam is created by inserting a thin scattering wire in the beam line while running a main experiment on a completely different physics case that takes place in the positron or neutron time-of-flight beam lines. Just one time-correlated electron per bunch is then guided to the direct electron beam station. The scattered out electrons have a negligible effect on the main experiment. Thus, parasitic mode allows the use of ELBE for detector tests without devoting primary beam time to the experiment. 

A common feature of both modes of operation is that most bunches are actually empty, i.e. devoid of electrons, with an effective electron beam rate that can vary between 10$^0$-10$^5$ s$^{-1}$.

\begin{figure*}[tb]
    \centering
    \includegraphics[width=0.950\textwidth]{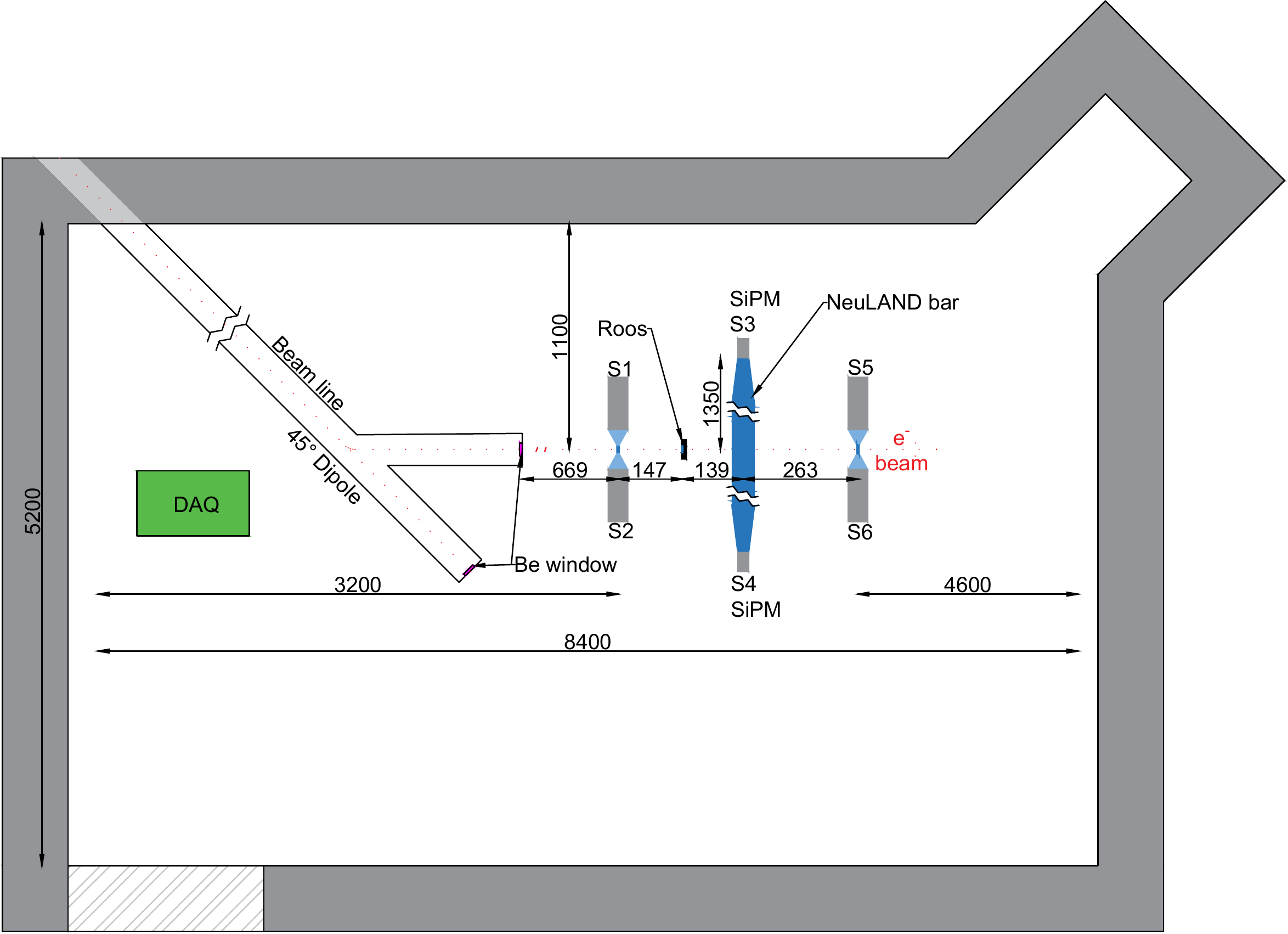}
    \caption{Schematic top view of the experimental setup at ELBE. The beam enters from the top left of the figure. Dimensions in mm. The beam line is 1.4\,m above the floor, and the room has a total height of 3.6\,m.}
    \label{fig:ELBE_plan}
\end{figure*}

\subsection{Setup for in-beam studies at ELBE}
\label{subsec:Electronbeamsetup}

The direct electron beam setup is mounted in an experimental cave that is shielded by 1.5 m thick walls of borated concrete (\autoref{fig:ELBE_plan}).
There are two beryllium exit windows located, respectively, at the 0$^\circ$ and 45$^\circ$ exits of a 45$^\circ$ deflection magnet. 

The experimental setup is placed immediately after the beryllium window at the 45$^\circ$ magnet exit. The NeuLAND bar under study is bracketed by two 5 mm thick plastic scintillators of $20\times20$\,mm$^2$ size for the definition of the electron beam and for excluding backscattered electrons. Each of these reference scintillators is each read out by two Philips XP2020 PMTs called S1, S2, S5, and S6, respectively. The time reference is provided by the electron gun RF signal instead. 
For an absolute dose calibration, an ionisation chamber is inserted between the first reference scintillator and the NeuLAND bar. The Roos-type ionisation chamber\footnote{PT34001 made by PTW Physikalisch-Technische Werkstätten Dr. Pychlau GmbH, Freiburg, Germany} has a circular sensitive area of 16\,mm diameter and 2\,mm thickness and is read out by a PTW Unidos dose rate meter placed in the counting room. 

The experiments reported here are limited to the central position of the NeuLAND bar.

\subsection{Off-board electronics and data acquisition}
\label{subsec:ExpElectronics}

The data acquisition relied on a completely digitiser-based system. 

Up to four high-speed cPCI Acqiris DC 282 cards were used in coupled mode. Each of these cards, in turn, includes four DC coupled 10\,bit channels with an input impedance of 50\,\textOmega\ and 0.05-5\,V input range at a maximum input bandwidth of 2\,GHz. The standard sampling rate is 2\,GS/s, and by combination of all channels of one card it is extendable to 8\,GS/s, especially for time resolution measurements. 

The waveform data are continuously recorded without a trigger and streamed to a PC with cPCI interface and a custom made control software for controlling the digitisers based on the Acqiris API. During the later, offline analysis, software triggers were defined as described below in the relevant sections.  

For most of the experiments reported here, the Acqiris system was placed in the experimental cave and controlled remotely from the counting room. The radiation dose was low and did not pose any danger to the electronics.

\subsection{Laser setup for off-beam calibrations}
\label{subsec:ExpPiLas}

As a first step, the SiPMs were illuminated with attenuated laser light from a picosecond diode laser unit (\mbox{PiLAS}\footnote{PiL042XSM by Advanced Laser Diode Systems, Berlin, Germany}) to record photon spectra where single fired pixels are visible. The 421\,nm laser wavelength was chosen to match the 420\,nm wavelength of the RP-408 scintillation light. The calibration was repeated for each of the SiPMs used, applying a list of different operating voltages near the operating voltage range recommended by the manufacturer.

For some of the laser calibrations, the on-board amplifier was not used and instead the individual SiPM signals were routed out. For those cases, the un-amplified signals passed through a CAEN N978 non-distorting amplifier, and afterwards corrected back in the offline calibration using the known gain of the CAEN N978. The signals amplified by the N978 were not used for timing purposes, only for the calibration of the charge per fired pixel. 

\subsection{Setup for off-beam dark count rate study}
\label{subsec:ExpDark}

For the measurement of the dark rate, the prototypes were removed from the respective NeuLAND bar. Each of the prototypes, in turn, was then placed in a lightproof box that was temperature stabilised at room temperature (20 $\pm$ 2 $^\circ$C). In order to avoid possible bias due to the trigger condition, the digitiser was triggered by an external 100\,kHz pulser, and after a trigger, waveform traces of 100\,\textmu s length were recorded. The time and pulse height of possible dark pulses were then extracted in the offline analysis.

\subsection{NeuLAND DAQ-based setup at GSI for cosmic-ray muon study}
\label{subsec:ExpGSI}

For the cosmic-ray study, two NeuLAND bars were placed one atop the other in the test station at GSI. One NeuLAND bar was instrumented on one side with the standard NeuLAND photomultiplier tube and on the other side with the present prototype, equipped with Hamamatsu S14160-6075 SiPMs. The second NeuLAND bar was instrumented with the standard PMTs on both sides.  

For these studies, the standard NeuLAND data acquisition with the FQT-TAMEX electronics has been used. It has been described in details previously \cite{Boretzky21-NIMA}. Briefly, FQT-TAMEX consists of a multi-hit charge-to-time converter using a linearized time-over-threshold method, TAMEX time-to-digital converter and is connected with a 2 Gb/s backplane to a PC which hosts several modules including the TRIXOR module that may be connected to the main R$^3$B trigger bus from the main data acquisition system. The time resolution of FQT-TAMEX is $\sigma_t \approx$ 8\,ps, low enough that it does not limit the overall time resolution of the system.

For the SiPM-instrumented channel, an FQT card with a resistive input was used instead of the usually installed transformer for the input of PMT signals. This card provided a better matching of the present prototype's voltage driven output.

\section{Data analysis procedures and calibrations}
\label{sec:DataAnalysis}

In the present section, the experimental procedures and analyses are presented. First, the analysis of the digitizer-based data is described (\autoref{subsec:Waveforms}). 

Subsequently, the charge calibration is described step by step: First, the observed integrated charge of the SiPM prototype output signal is converted to a number of fired SiPM pixels, using a PiLAS picosecond laser system (\autoref{subsec:AnalysisPiLas}).
Second, the fired SiPM pixels are converted to a number of minimum ionising particles passing the NeuLAND bar, using ELBE experiments (\autoref{subsec:AnalysisMIP}). 
Third, the in-beam data from the NeuLAND bar read out by SiPMs are calibrated against two types of ancillary in-beam detectors, namely reference scintillators read out by PMTs (\autoref{subsec:AnalysisS1S2}) and an ionization chamber (\autoref{subsec:AnalysisRoos}).

\begin{figure*}[t!!]
    \centering
    \includegraphics[width=\textwidth,trim=0 0 0 18mm,clip]{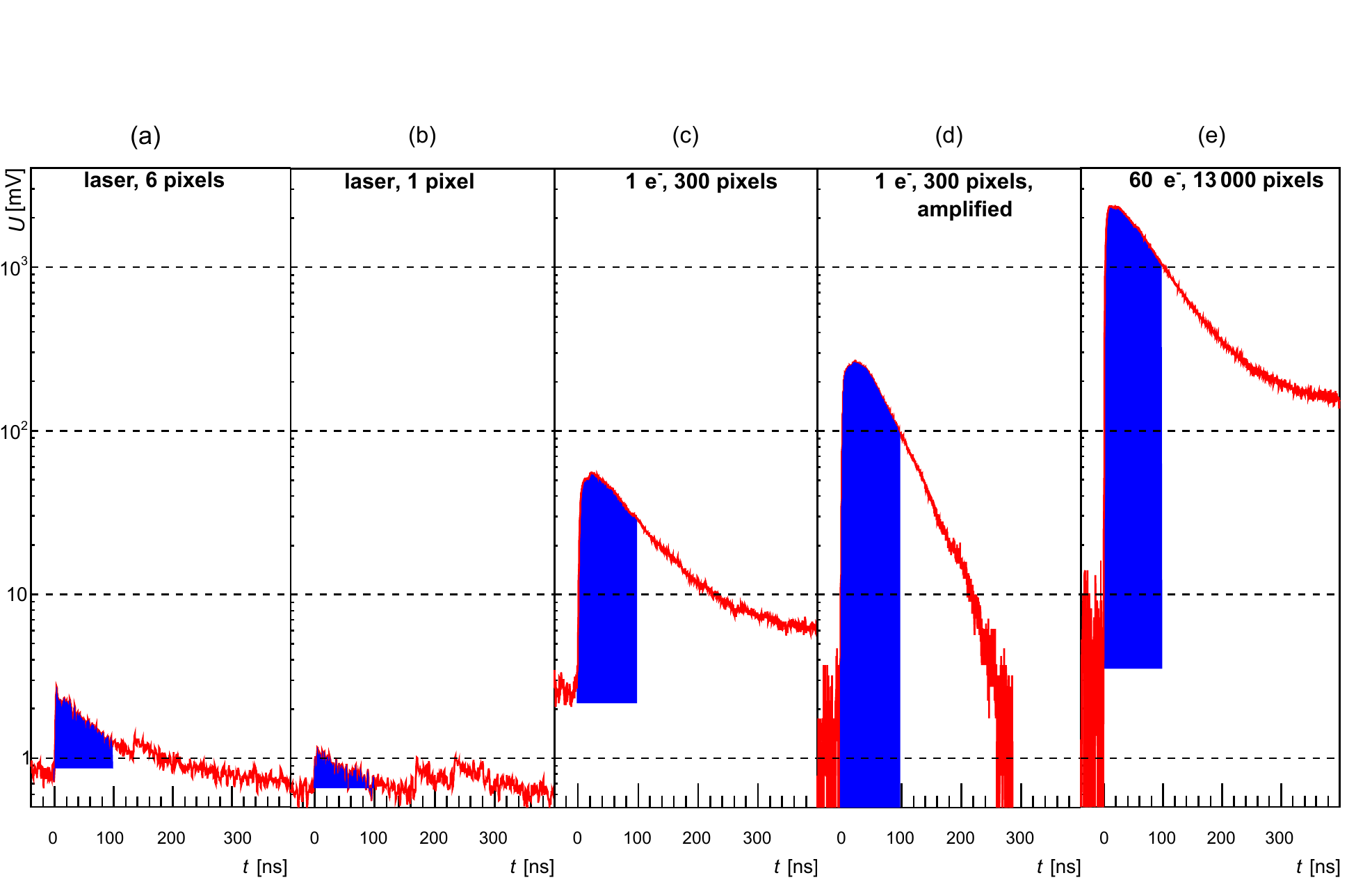}
    \caption{Typical waveforms produced by a Hamamatsu S14160-6075 SiPM. The rising edge time determined by the software CFD are aligned to $t=0$\,ns. -- (a) Laser measurement with 6 firing pixels and dark event afterwards. -- (b) Single firing pixel. -- (c) Single electron crossing NeuLAND bar (300 firing pixels). -- (d) Same as (c) but integrated preamplifier used and signal shaping clearly visible. -- (e) SiPM output for a bunch of 60 electrons hitting the bar (13000 firing pixels). The blue area marks the charge integration window of 100\,ns starting at the CFD trigger. Note that only signal (d) is amplified.}
    \label{fig:waveforms}
\end{figure*}

\subsection{Determination of the integrated charge from the recorded waveforms}
\label{subsec:Waveforms}

The recorded waveform data are analysed offline, using a digital constant fraction digitizer (CFD) technique for the time definition of the rising edge, with a typical CFD fraction of 0.12 applied after baseline subtraction. The unamplified signals decay by a factor 10 within 300\,ns, but the subsequent, full return to the baseline is much slower, on the order of 1\,\textmu s. The amplified signals fully return to the baseline with a time constant of 50\,ns (\autoref{fig:waveforms}).

The measured voltage is converted to a current using the Acquiris system's input impedance of 50\,$\Omega$. 
The current is then integrated for an integration time window of 100\,ns (blue shaded area in \autoref{fig:waveforms}), starting from the rising edge determined by the software CFD. The window length of 100\,ns was adopted as a compromise that works both for unamplified signals and for amplified signals. This window length mitigates random noise (dominant at very short integration times) and effects of recharging (dominant at very long integration times). The resulting integrated charges are filled in a charge spectrum.

\begin{figure}[bt]   \centering
    \includegraphics[width=\columnwidth,clip,trim=6mm 3mm 20mm 5mm]{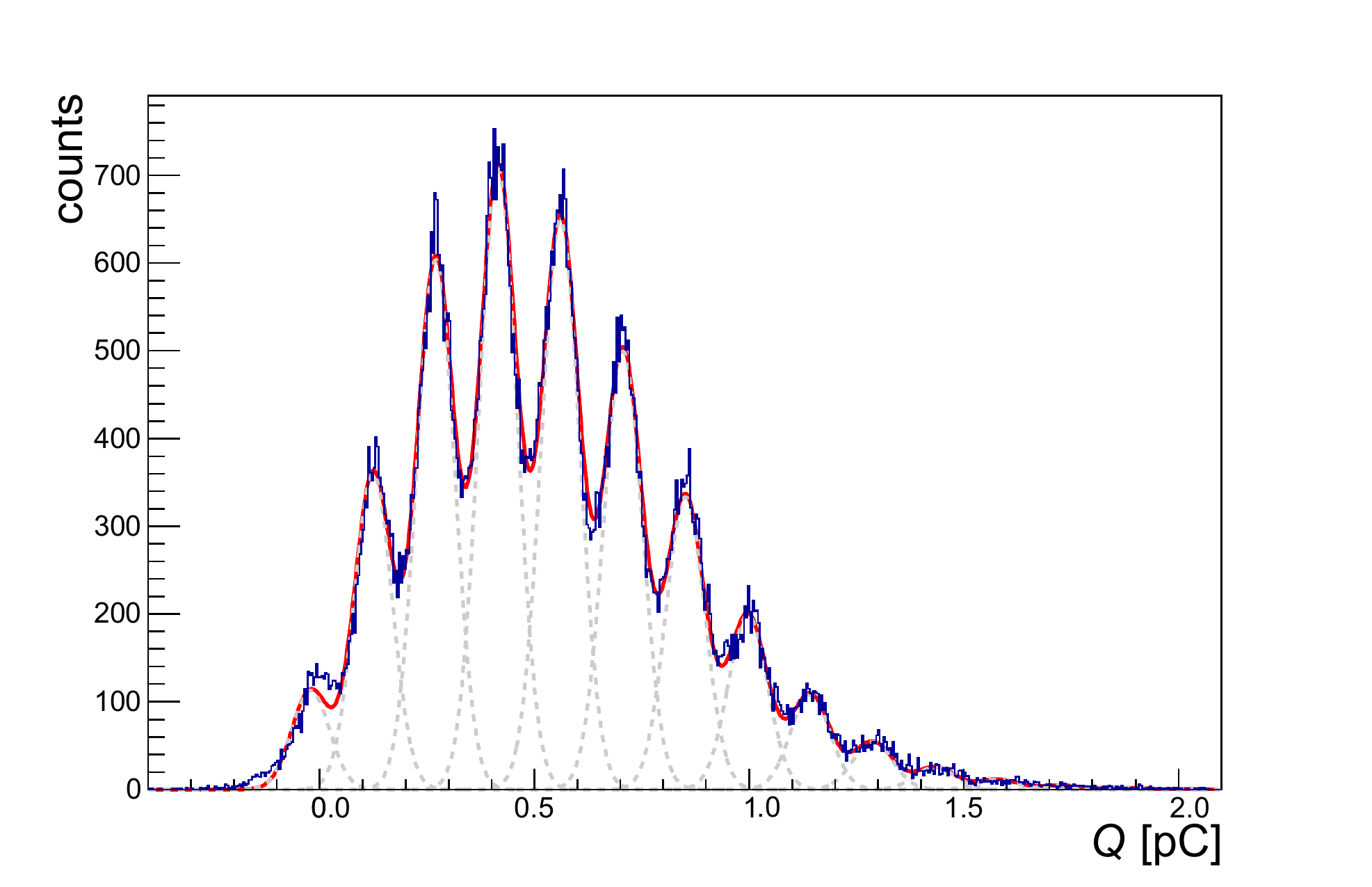}
    \caption{Laser-based calibration of a single Hamamatsu S14160-6075 SiPM at $U_{\rm OP}$= 40\,V, with 100\,ns integration time. Fits to the peak positions for 1-10 fired SiPM pixels (after the pedestal) are shown in red. The fits served to determine the output charge $Q_{{\rm single}, \, i}$ per firing SiPM pixel. See text for details.}
    \label{fig:singles2}
\end{figure}

\subsection{Laser-based determination of the number of fired pixels from the integrated charge}
\label{subsec:AnalysisPiLas}

The integrated charge spectra from the PiLas laser runs clearly show the signatures of individual firing pixels. Such laser-based spectra have been taken for each combination of prototype studied and operating voltage. Figure \ref{fig:singles2} shows a typical example.

From these individual spectra (e.g. \autoref{fig:singles2}), for a given combination $i$ of prototype and overvoltage, the charge per single firing SiPM pixel $Q_{\rm single, \, i}$ has been determined using the averaged offset between peaks after a fit. For the five prototypes studied here, values of $Q_{{\rm single}, \, i}$ = 0.1-0.5\,pC have been found for voltages on the efficiency plateau. 

Using the PiLas-obtained $Q_{{\rm single}, \, i}$ value, for any given SiPM $i$ at the same operating voltage and charge integration time, the number of fired SiPM pixels $n_{\rm fired}$ is derived from the observed charge $Q_i$ using the relation
\begin{equation}
    n_{\rm fired, \, i} = \frac{Q_i}{Q_{{\rm single}, \, i}}
    \label{eq:nfired}
\end{equation}
Using this calibration, the charge spectrum is then converted to a spectrum of fired SiPM pixels $n_{\rm fired}$. 
%


\subsection{In-beam determination of the number of minimum ionising particles from the number of fired pixels}
\label{subsec:AnalysisMIP}

As a next step, in-beam experiments at ELBE have been conducted in order to experimentally determine the number of fired pixels for one minimum ionising particle (35 MeV electron) passing the NeuLAND bar near its center. The energy loss of such particles is about 10\,MeV in the 5\,cm thick NeuLAND bar. 

For this calibration, the geometrical average of the numbers of fired pixels for the two SiPM-based prototypes S3 and S4 instrumenting the two sides of the NeuLAND bar, here called $n_{\rm fired, \, 3}$ and $n_{\rm fired, \, 4}$, is taken, in order to compensate attenuation along the length of the NeuLAND bar. 
The numbers of fired pixels, in turn, are obtained from the laser-based calibration of each side as described above (\autoref{eq:nfired}).

\begin{figure}
     \centering
        \includegraphics[width=\columnwidth,trim=12mm 9mm 11mm 12mm]{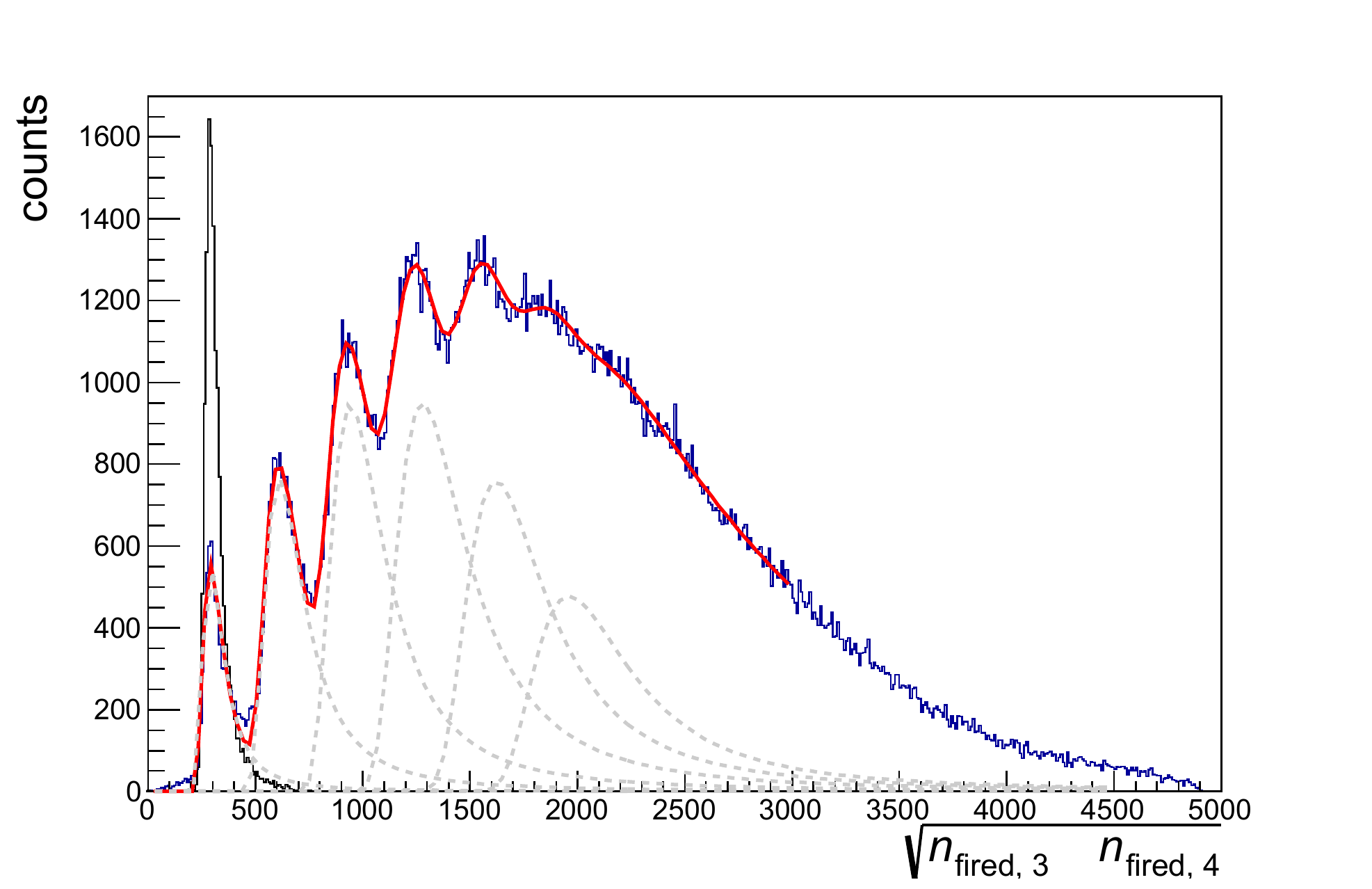}
    \caption{%
    Spectrum of the geometrically averaged number of fired pixels $\sqrt{n_{\rm fired, \, 3} \times n_{\rm fired, \, 4}}$ for parasitic mode (black spectrum) and for single-user mode with a gate voltage of $U_{\rm G} = 9.5$\,V (blue spectrum). The data have been taken with an assembly of four {\tt onsemi} J Series 60035 SiPMs at 29\,V.
    The red fitted curve shows the sum of individual Landau distributions (dashed curves) fitting the single-user mode spectrum. See text for details.
    }
    \label{fig:PHS_singlepeaks} \label{fig:PHS_epeaks}
\end{figure}

The observed pulse height spectrum can be fitted by a single Landau distribution for parasitic mode and by a sum of Landau distributions for single-user mode (\autoref{fig:PHS_singlepeaks}). A typical example is given by the {\tt onsemi} J Series 60035 SiPM. For the prototype using this SiPM type, the observed Landau distributions have a mean value of $\mu_{\rm J35} = 331$, and they are offset from each other by the same value $\mu_{\rm J35}$ (\autoref{fig:PHS_singlepeaks}). This fitted value $\mu$ is the number of fired pixels per 35 MeV electron passing the NeuLAND bar. 

For the other prototypes, when operated on their respective efficiency plateaus, values of $\mu$ = 230-400 fired pixels per 35 MeV electron are found.

As a result, the number of minimum ionising particles (mip) passing the NeuLAND bar is obtained by 
\begin{eqnarray}
    n_{\rm mip}^{\rm obs} & = & \frac{\sqrt{n_{\rm fired, \, 3} \times n_{\rm fired, \, 4}}}{\mu} \nonumber \\
    & = & \frac{1}{\mu} \left[ \frac{Q_3}{Q_{\rm single, \, 3}} \times \frac{Q_4}{Q_{\rm single, \, 4}} \right]^\frac{1}{2}
    \label{eq:neobs}
\end{eqnarray}

 The measured values of $\mu$=230-400 per 35 MeV electron depositing 10 MeV of energy in the NeuLAND bar correspond to 23-40 photoelectrons (p.e.) per MeV$_{ee}$ (MeV deposited energy). The four SiPMs used in one prototype cover only 30\% of the 490\,mm$^2$ surface of the light guide 
 at the bar end. For comparison, a standard 1" cylindrical PMT covers this entire area. This 
 inefficiency of coverage is partially compensated by approximately 2 times higher photon detection 
 efficiency of SiPM compared to PMT.

\subsection{Reference scintillator S1S2 calibration}
\label{subsec:AnalysisS1S2}

In order to enable the study of possible saturation and offset effects, it is useful to also determine the number of minimum ionising particles passing the NeuLAND bar, but without using the NeuLAND bar data. Two methods are used to study this effects for the single-user experiments and are described in this and the following subsections.  

First, the number of 35\,MeV electrons passing the NeuLAND bar is determined based on an external reference scintillator. This 0.5\,cm thick initial scintillator is called S1S2 and has an area of 2$\times$2 cm$^2$. It is placed in the path of the electron beam in front of the NeuLAND bar (\autoref{fig:ELBE_plan}). 

For the purposes of the calibration, the observed charge per bunch in S1S2, as given by the geometric average $    Q_{12} = \sqrt{Q_1 \times Q_2}$ of the charges  $Q_{1,2}$ of the two photomultipliers reading out the two sides has been measured 
for single-user beam at gate voltages ranging from 8.5-11.5\,V. 

For each run, in addition to $Q_{12}$, also the number of electrons per bunch in the SiPM-readout NeuLAND bar $n_{\rm mip}^{\rm obs}$ is determined. Note that at the lowest gate voltage studied, $U_{\rm G}$ = 8.5\,V, the average number of electrons per bunch is below 1, because there are some triggered events where the signal in the NeuLAND bar is very low, presumably due to electrons that are scattered out of the beam or due to false triggers from background electrons.

At low values of $Q_{12}$, the data show strict proportionality between $n_{\rm mip}$ and $Q_{12}$ (\autoref{fig:Synopsis}, full red curve):
\begin{equation}
    n_{mip} = 0.095\, {\rm pC}^{-1} \times Q_{12} 
    \label{eq:Q12}
\end{equation}
However, for $Q_{12} >$ 100\,pC (corresponding to $n_{\rm mip}^{\rm obs} >$ 10) some deviation from linearity is observed  (\autoref{fig:Synopsis}, dashed red curve). Such a deviation may in principle be due to the saturation behaviour of the SiPM.

\subsection{Roos ionization chamber calibration}
\label{subsec:AnalysisRoos}

For the Roos ionisation chamber, only integrated charges over times of typically a few minutes are available. 
The main precondition needed to use the Roos chamber in an absolutely calibrated mode is only partially fulfilled. Namely, the size of the beam is Gaussian with $\sigma \approx 1$\,cm \cite{Yakorev11-NIMA}, significantly larger than the Roos chamber radius of 0.78\,cm but still below the minimum field size of 4$\times$4\,cm$^2$ recommended in the Roos chamber datasheet.

In addition to this caveat, another correction is needed because the Roos chamber used is significantly smaller than the 5\,cm wide NeuLAND bar under study. A GEANT4 simulation of the electron beam propagating through the Roos chamber and the NeuLAND bar showed that only $f_{\rm GEANT} = 0.3\pm0.1$  of the incoming electrons hitting the NeuLAND bar also passed the smaller Roos chamber. The uncertainty on  $f_{\rm GEANT}$ is given by the 30\% error on the width of the effective beam profile ($\sigma \approx$ 1\,cm \cite{Yakorev11-NIMA}) and by the not precisely known thickness of the beryllium exit window used to couple out the beam to air. 

Taking these effects into account and neglecting for the moment any background in the Roos chamber, a theoretical calibration factor $f_{\rm Roos}^{\rm theory}$ of 
\begin{eqnarray}
    f_{\rm Roos}^{\rm theory} & = & \frac{N_{\rm D, W} \times k_{\rm Q}}{\frac{dE}{dx} \times \frac{1}{A}  \times f_{\rm GEANT}} = \nonumber \\
     & = & \frac{8.253 \times 10^7 \frac{\rm Gy}{\rm C} \times 0.88}{2.88 \frac{\rm MeV \, cm^2}{\rm g} \times \frac{1}{1.91 {\rm cm}^2}  \times 0.3} = \nonumber \\
    & = & 1.0^{+0.5}_{-0.3} \, {\rm aC}^{-1}
    \label{eq:RoosTheory}
\end{eqnarray}
is obtained. Here, $N_{\rm D, W} = 8.253 \times 10^7$ Gy/C is the $^{60}$Co-based water energy dose calibration value of the Roos chamber, $k_{\rm Q} = 0.88$ is the beam quality correction needed to apply $N_{\rm D, W}$ to 35 MeV electrons, $\frac{dE}{dx}$ is the energy loss of 35\,MeV electrons in water, and $A=1.91$\,cm$^2$ is the area of the Roos chamber. Then, the theoretically predicted number of electrons traversing the chamber $n_{\rm mip}$ is given by 
\begin{equation}
    n_{\rm mip}  =  f_{\rm Roos}^{\rm theory} \times Q_{\rm Roos}
\end{equation}

In order to use the data of the Roos chamber quantitatively, in addition to these theoretical considerations, an experimental calibration was performed. The electron beam was used in single-user mode, with 2 kHz signal rate. The Roos current was averaged over a time of 2-5 minutes with stable accelerator parameters, and converted to a charge $Q_{\rm Roos}$ by multiplying with the 0.5\,ms integration window size given by the 2\,kHz beam rate. The data are shown in \autoref{fig:Synopsis}, blue triangles.

The Roos data are well described by a linear curve with a constant background of 1.6\,aC (integrated current over 0.2\,ms).
This Roos chamber background may be explained by two factors: First, activation of the room and beam line, and second, so-called dark electrons created by discharges and then accelerated in the outer areas of the beam line. The latter are not correlated in time to the RF signal of the electron gun. The background value of 1.6 aC is fitted as a free parameter in the calibration curve and lies in the range of values obtained during dedicated room background measurements undertaken during beam interruptions.

Using now the above described calibration for the reference charge $Q_{12}$ (\autoref{eq:Q12}), the following relation for the experimentally observed 
calibration factor $f_{\rm Roos}^{\rm exp}$
is obtained:
\begin{eqnarray}
    f_{\rm Roos}^{\rm exp} & = & \frac{n_{\rm mip}^{\rm obs}}{Q_{\rm Roos}^{\rm no\,bg}} \nonumber \\
        & = & \frac{0.095 \, {\rm pC}^{-1} \times Q_{\rm 12}}{Q_{\rm Roos}^{\rm no\,bg}} \nonumber \\
        & = & \frac{0.095 \, {\rm pC}^{-1}}{Q_{\rm Roos}^{\rm no\,bg}} \times \frac{Q_{\rm Roos}^{\rm no\,bg}}{0.042 \, {\rm aC / pC}} \nonumber \\
        & = & (2.3 \pm 0.2) \, {\rm aC}^{-1}
    \label{eq:RoosExperiment}
\end{eqnarray}
where 
\begin{equation}
    n_{\rm mip}^{\rm obs}  =  f_{\rm Roos}^{\rm exp} \times Q_{\rm Roos}^{\rm no\,bg}
\end{equation}

The experimental calibration factor (\autoref{eq:RoosExperiment}) is in fair agreement with the theoretical one (\autoref{eq:RoosTheory}). In the following, the experimental calibration will be used.

\subsection{Discussion of the two S1S2 calibrations}

\begin{figure}[t]
     \centering
         \includegraphics[width=\columnwidth]{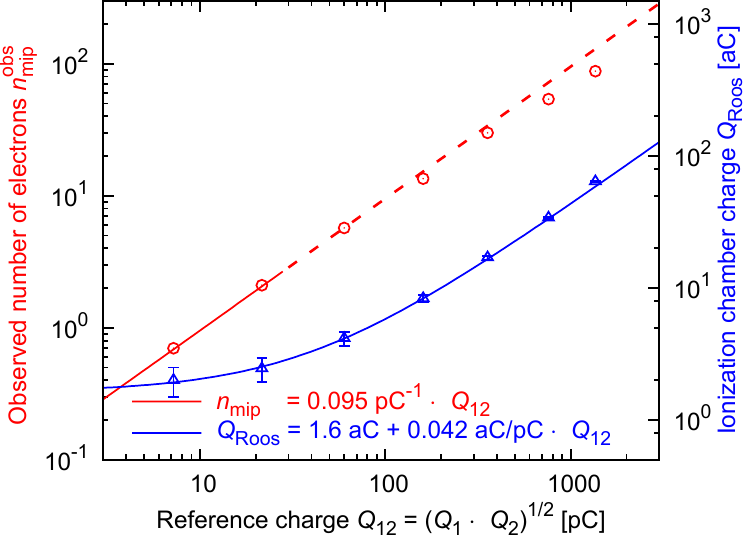}
    \caption{Synopsis of the electron-beam calibration data (sections \ref{subsec:AnalysisS1S2} and \ref{subsec:AnalysisRoos}), together with fitted calibration curves, Equations (\ref{eq:Q12}) and (\ref{eq:RoosExperiment}). }
    \label{fig:Synopsis}
\end{figure}

The two calibrations described in \autoref{subsec:AnalysisS1S2} and \autoref{subsec:AnalysisRoos} are now discussed together with the aid of \autoref{fig:Synopsis}. 

At low values of the reference charge $Q_{12}$, hence low observed number of electrons per bunch $n_{\rm mip}^{\rm obs}$, there is a strict proportionality between these two quantities, hence the reference charge can be used as proxy for the observed number of electrons per bunch. At higher values of $Q_{12}$, $n_{\rm mip}^{\rm obs}$ is lower than predicted by linearity, which is qualitatively in agreement with the onset of saturation behaviour in the SiPMs used to determine $n_{\rm mip}^{\rm obs}$.

At the same high values of $Q_{12}$, however, $Q_{12}$ is strictly proportional to the Roos chamber charge $Q_{\rm Roos}$. At these values, the necessary background correction of the Roos chamber (see above, \autoref{subsec:AnalysisRoos}) introduces only negligible uncertainty. 

Importantly, the Roos chamber has been designed to show a strictly linear response up to doses of Gy, orders of magnitude above the range explored here. The calibration of S1S2 against the Roos chamber data (Figure \ref{fig:Synopsis}, blue curve) shows that there are no saturation effects in  the $Q_{12}$ reference charge. Therefore, in the intensity range used here, S1S2 can be used as a secondary linearity standard, derived from the well-known Roos chamber linearity.

Summarising, the data of the two ancillary detectors used here show that $Q_{12}$ is a suitable proxy for the true number of electrons per bunch over the entire intensity range explored here.

\section{Results of the in-beam studies at ELBE} 
\label{sec:ResultsELBE}

In this section, the results of the experiments using the direct electron beam coupled out into air are shown. 

In \autoref{subsec:EfficiencyTimeResolution}, the efficiency and time resolution of a NeuLAND bar instrumented with two identical SiPM-based prototypes are determined using the parasitic-mode electron beam with just one electron per bunch. These studies aim to meet or exceed previous ELBE-based results on these two parameters \cite{Reinhardt16-NIMA}. 

Subsection \ref{subsec:Saturation}, instead, uses the single-user mode electron beam with its scaleable bunch charge corresponding to 1-30 electrons per bunch, in order to study output linearity and saturation effects.

\subsection{In-beam determination of efficiency and time resolution} 
\label{subsec:EfficiencyTimeResolution}

\begin{figure*}[tb]
    \centering
    \includegraphics[trim={0 3.5cm 0 0}, width=0.8\linewidth]{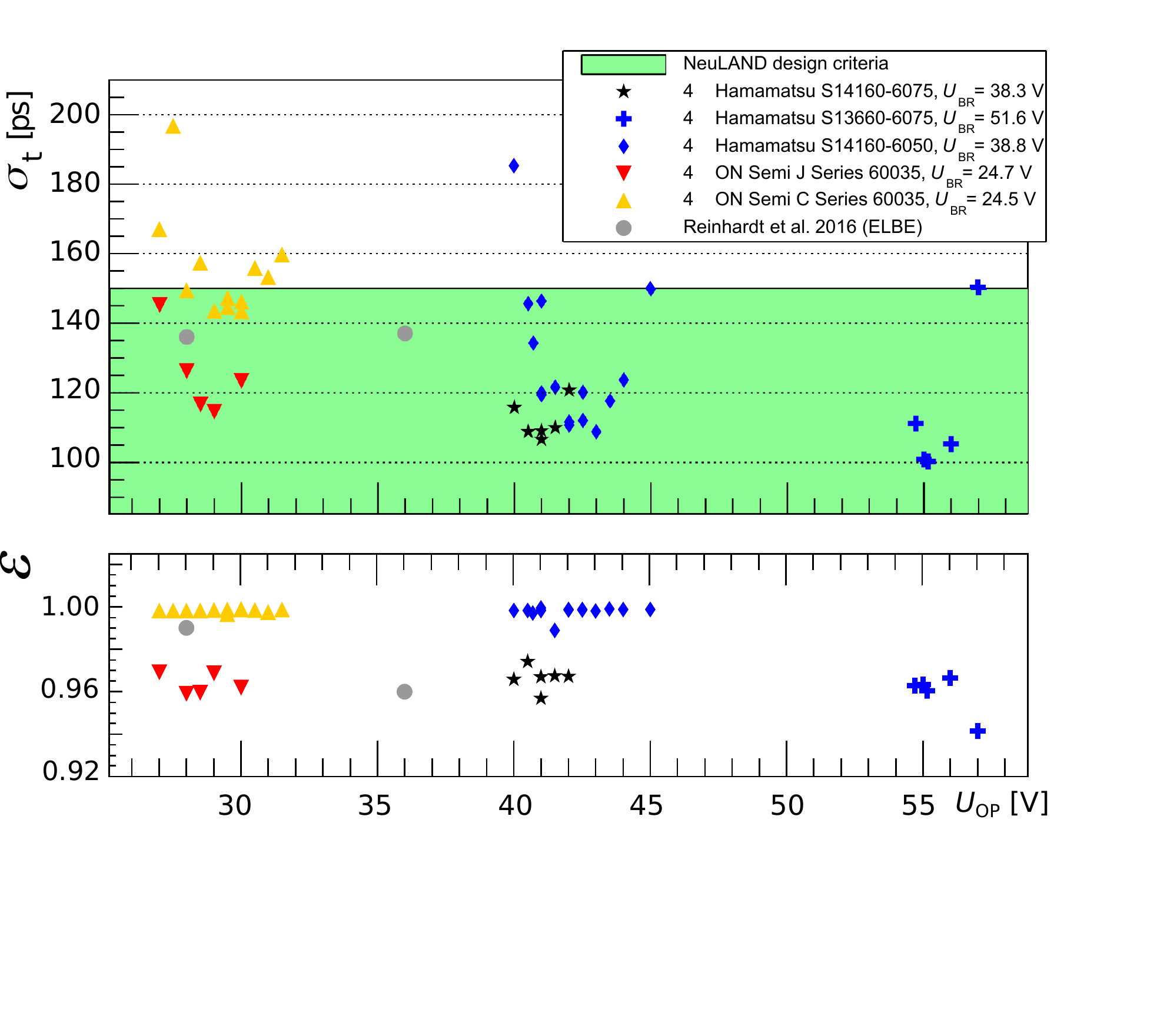}
    \caption{Time resolution $\sigma_{t}$ and efficiency $\epsilon$, see Equation (\ref{eq:Efficiency}), as a function of operating voltage $U_{\rm OP}$. All data have been taken with the NeuLand bar, instrumented with SiPM arrays at both ends, at ELBE. All investigated SiPM models meet and exceed the NeuLAND design criteria in a specific operating voltage range. For comparison, arrays of SensL C-series and FBK NUV SiPMs investigated previously are also shown (Reinhardt {\it et al.} 2016, \cite{Reinhardt16-NIMA}).}
    \label{fig:sigma_vs_vop}
\end{figure*}

Using the single-electron beam, the time resolution and efficiency were determined as function of the operating voltage for all the SiPM types studied here. 

The time resolution $\sigma_t$ was determined in the same way as in previous work \cite{Reinhardt16-NIMA}. Briefly, the respective trigger times obtained by the software CFDs for prototypes at both ends of the bar were averaged in order to remove the effect of the lateral size of the electron beam, and the reference time given by the accelerator RF signal was then subtracted event by event from this average. The resultant distribution was almost free of time-walk effects and simply fitted by a Gaussian to obtain  $\sigma_t$. 

In order to be conservative, in this work the contributions of the DAQ system and the time reference from the accelerator, both below 23\,ps, were not subtracted from the $\sigma_t$ values. The time resolutions shown here have a typical uncertainty of 5\,ps, which is given by the variation of measured time resolutions in different experimental runs taken at different times, i.e. the run-to-run reproducibility.

The efficiency $\epsilon$ is defined as the ratio between the number of NeuLAND triggers $N((\rm S3 \wedge S4)  \wedge (S1  \wedge S2 \wedge S5 \wedge S6))$, on the one hand, and the rate of reference triggers $N(\rm (S1  \wedge S2 \wedge S5 \wedge S6)$, on the other hand:
\begin{equation}\label{eq:Efficiency}
    \epsilon = \frac{N((\rm S3 \wedge S4)  \wedge (S1  \wedge S2 \wedge S5 \wedge S6))}{N(\rm (S1  \wedge S2 \wedge S5 \wedge S6)}
\end{equation}

When comparing different SiPM types, it is observed that an increase in pixel size improves the time resolution. This is due to the higher fill factor, hence higher photon detection efficiency (PDE), of devices with larger pixels and less insensitive interpixel areas.  

For all SiPMs studied, the data show a range in operating voltages around 2-3\,V overvoltage that is several volts wide where the NeuLAND timing goal of $\sigma_t \leq 150$\,ps (green area in \autoref{fig:sigma_vs_vop}) and the NeuLAND efficiency goal of $\epsilon \geq95$\% are both fulfilled. The lower edge of this optimal operating range is given by the decrease in PDE for lower overvoltages. 

The upper edge, instead, is given by the onset of a significant dark rate. The degradation of the time resolution may be cause by one of two effects, or a combination of both: First, a dark count rate that affects the baseline and, hence, the time resolution (e.g. Figure \ref{fig:waveforms}, panels (a) and (b)). Second, dark current that may affect the amplifier \cite{2022NIMPA103466745S}.

The present plateaus are sufficiently wide for the present purposes. In the literature, using a different electronics layout with a faster return to baseline than in the present case, a similar level ($\sigma \sim$ 140\,ps) of the timing plateau has been reported, but with a significantly wider, up to 4-6 V overvoltage, plateau width  \cite{2022JInst..17P1016K}.

\subsection{In-beam study of linearity and saturation effects} 
\label{subsec:Saturation}

Saturation in the SiPMs will occur mainly because of the finite number of pixels $n_{\text{pixels}}$ in each SiPM. If a pixel has already fired and a second photon is hitting the pixel in the same time, it cannot fire again. So the observed number of fired pixels $n_{\text{fired}}$ is reduced below the expected number of firing pixels $n_{\text{seed}}$. This effect can be described with a statistically motivated approach \cite{Gruber14-NIMA}:
\begin{equation} \label{eq:Saturation}
 n_{\text{fired}}^{\rm obs}  =  n_{\text{pixels}} \left[ 1 - \exp\left( - \frac{n_{\rm seed}}{n_{\rm pixels}} \right)  \right]
\end{equation}
For practical purposes, here the linearity is quantified by its slope, which is given by the derivative of \autoref{eq:Saturation}:
\begin{equation}
 \frac{d n_{\text{fired}}^{\rm obs}}{d n_{\text{seed}}} = \exp\left( - \frac{n_{\rm seed}}{n_{\rm pixels}} \right) \label{eq:DeltaSaturation}
\end{equation}
In an ideal case, the obtained slope (\autoref{eq:DeltaSaturation}) should be close to unity to allow an accurate energy determination.

In the literature, studies of the SiPM saturation behaviour have been reported, e.g., illuminating the SiPM directly with laser light, either at the SiPM sensitive wavelength \cite{Weitzel19-NIMA}, or at shorter wavelength to cause fluorescence \cite{Tsuji20-JINST}. Here, a different approach is chosen. The number of ionising particles per bunch that are simultaneously crossing the NeuLAND scintillator bar is varied in ELBE single-user mode. In this way, the possible recharging of fired SiPM pixels (not visible when using precisely time-correlated laser photons \cite{Weitzel19-NIMA}) should be directly included in the results obtained. The same is expected to be true for saturation effects in the scintillation process itself (not visible when using lasers \cite{Weitzel19-NIMA,Tsuji20-JINST}).

\begin{figure*}[t]
    \centering
    \includegraphics[width=\textwidth]{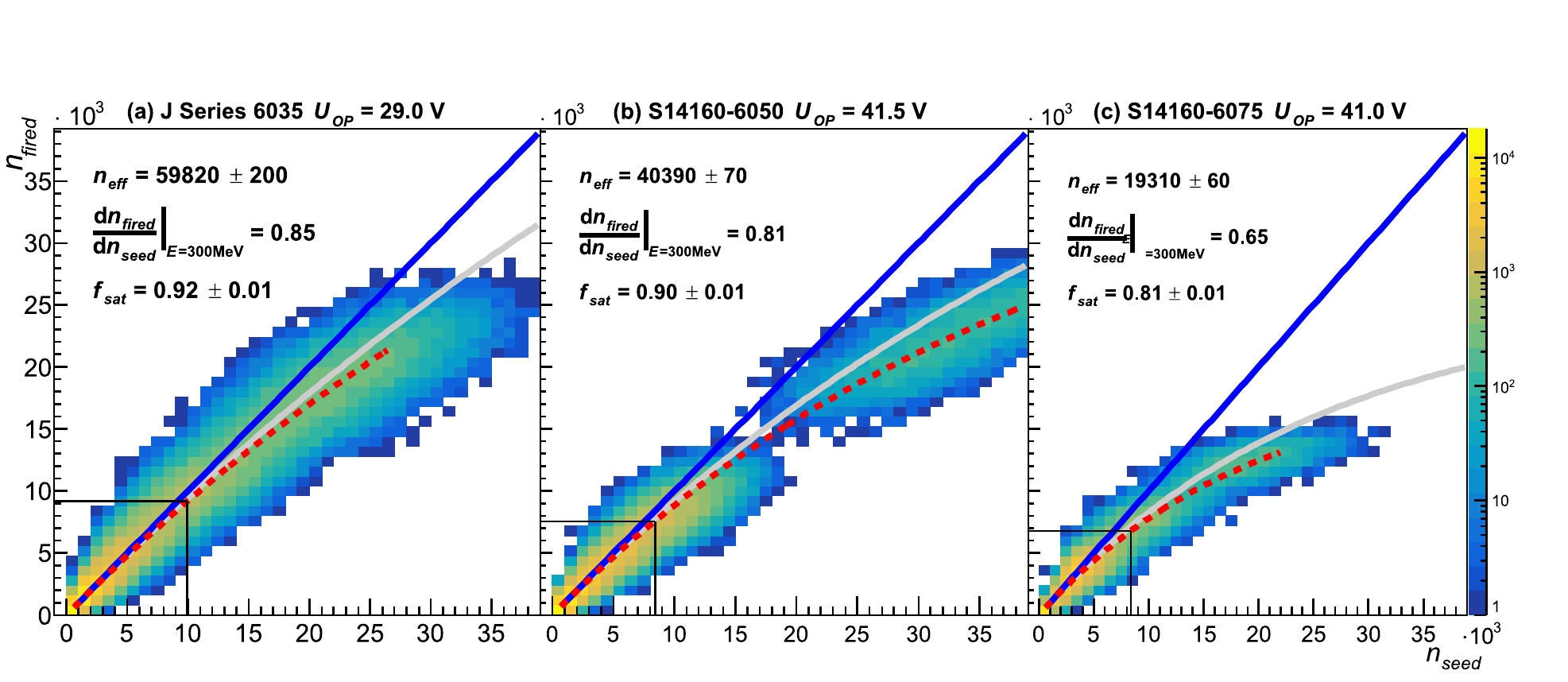}
    \caption{Number of fired pixels $n_{\text{fired}}$ as a function of the expected number of firing pixels $n_{\text{seed}}$ (from Equation (\ref{eq:Q12})) for three different SiPM types, at their respective optimal operating voltages. The blue and grey lines, respectively, show the expected behaviours without saturation and with the predicted saturation (\autoref{eq:Saturation}). The red dotted line is a fit to the data with $n_{\rm eff}$ as fit parameter. The black rectangle corresponds to an energy deposition of 300\,MeV, the expected maximum energy deposition in one NeuLAND bar. See text for details. }
    \label{fig:sat_SiPM}
\end{figure*}

The expected range of energy depositions in NeuLAND reaches up to 300\,MeV \cite{Boretzky21-NIMA}. For 10\,MeV energy deposition per 35\,MeV electron passing the NeuLAND bar, this corresponds to a range of 1-33 electrons per bunch. 

For this study, data in single-user mode were taken at several different gate voltages, corresponding to different average numbers of electrons per bunch. The data from different runs were accumulated to give an overall picture and fitted to \autoref{eq:Saturation}. For the fit, the number of pixels from the datasheet $n_{\rm pixels}$ was replaced by an effective number of pixels $n_{\text{eff}}$, which was fitted as a free parameter. The accumulated data and fitted curves are shown in \autoref{fig:sat_SiPM}.

The fitted effective number of pixels is generally somewhat lower than the datasheet value. For example, for the {\tt onsemi} J Series 60035 based prototype, the fit results in $n_{\text{eff}}$ = 59800$\pm$200, 35\% lower than the data sheet value of $n_{\text{eff}}$ = 4$\times$22292 = 89168. This corresponds to a stronger saturation effect than theoretically predicted. This is probably due to incomplete integration of large pulses due to the fixed integration window length (Figure \ref{fig:waveforms}) in our test environment. Another possible explanation is a reduction of voltage, hence detection efficiency and gain, for strong light pulses with a high number of firing pixels and a correspondingly high signal current \cite{vanDam10-IEEE}. 

For the NeuLAND application, the ratio $f_{\rm sat} = n_{\text{fired}}^{\rm obs} / n_{\rm seed}$ corresponding to the integrated saturation correction has been determined at a number of fired pixels equivalent to 300 MeV deposited energy. The determined values range between 0.81 and 0.92. With saturation corrections that are so close to unity, the calorimetric properties of the NeuLAND bar are still maintained, so no  effort was made to further decrease saturation effects. As expected, both the integrated saturation correction and its slope are largest for the largest pixel size (75 \textmu m), i.e. the lowest number of pixels per SiPM.

\section{Results of the off-beam studies at HZDR and at GSI}
\label{sec:ResultsOffline}

\subsection{Off-beam study of the dark rate} 
\label{subsec:Darkrate}

Dark counts generated by thermal noise electrons may cause an avalanche in a pixel. The frequency of occurrence of this effect depends on parameters like the overvoltage, temperature, radiation damage, and the production process in general \cite{Engelmann18-EPJC, Engelmann18-PhDTh}. In case of a high rate of such events, stochastic coincidences and cross-talk between pixels may lead to dark events with macroscopic amplitudes that resemble real scintillation events. 

In the case of the NeuLAND detector, such events can typically be rejected, because the dark signals are not time correlated to the start signal of the time of flight system. Still, a high dark count rate might overload the data acquisition or trigger generation system. 

\begin{figure}[t]
    \centering
    \includegraphics[width=0.45\textwidth]{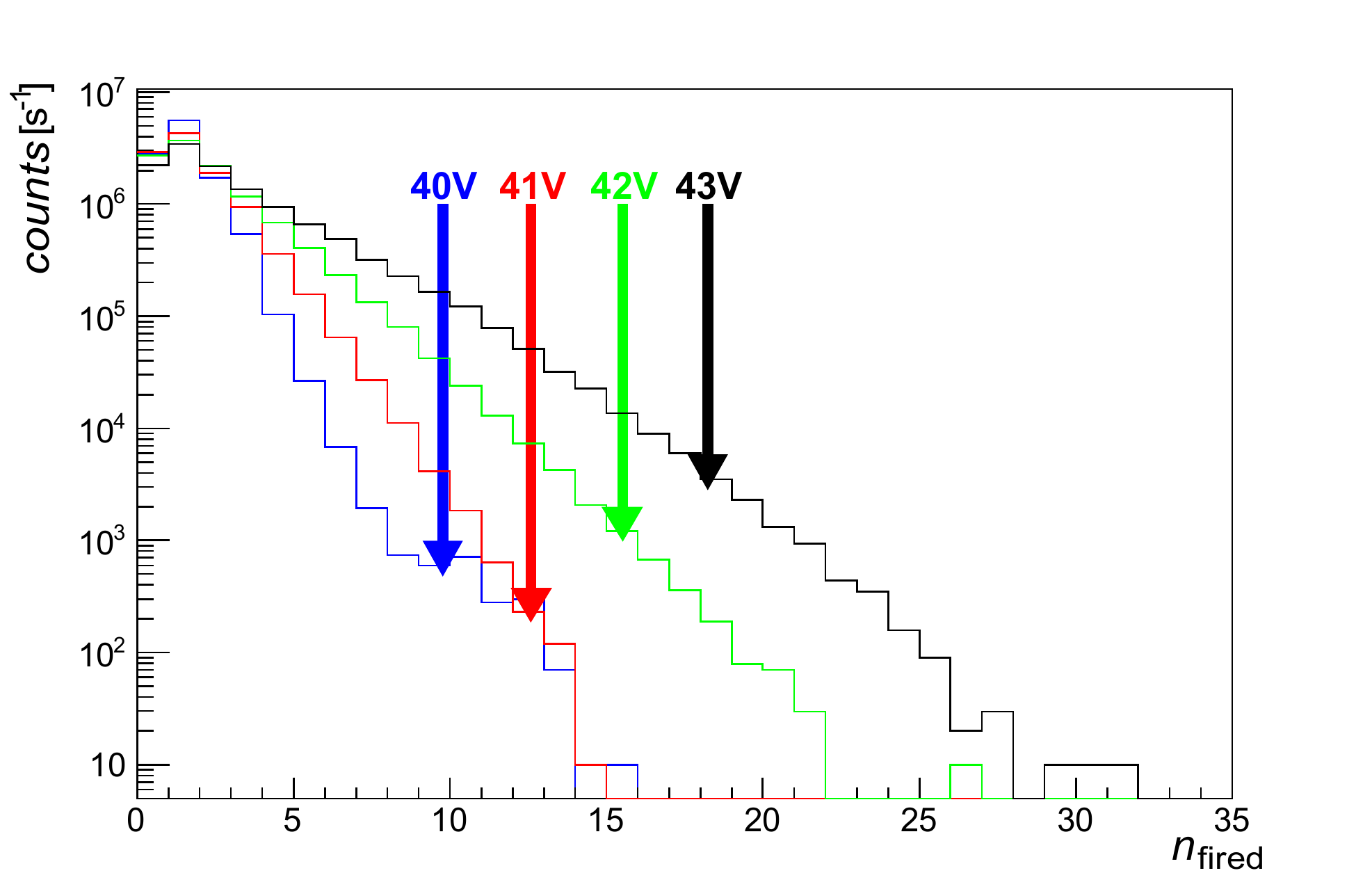}
    \caption{Dark spectra of a Hamamatsu S14160-6075 based prototype at room temperature (20\,$^\circ$C) at different operating voltages. Vertical arrows mark the start point where the NeuLAND trigger would fire, 0.5\,MeV$_{ee}$.}
    \label{fig:darks}
\end{figure}

The NeuLAND trigger is composed of so-called double plane triggers, where each double plane is made of two single planes of 50 NeuLAND bars each.  A double plane (dp) trigger is produced when any two of the $n=200$ photosensors of the dp show a signal of $\geq$0.5\,MeV$_{ee}$ in a coincidence time window of $\tau = 60$\,ns. The total dark rate $R_{\text{dp}}$ of a double plane by random coincidences can then be estimated using the measured dark rate $R_{\rm trig}$ of a single photosensor assembly, again taken for $\geq$0.5\,MeV$_{ee}$  signals.

\begin{eqnarray} \label{eq:darks}
R_{\text{dp}}  =  n \cdot (n-1) \cdot \tau \cdot R_{\rm trig}^2.
\end{eqnarray}

The measured dark count spectrum (\autoref{fig:darks}, for the typical example of the Hamamatsu S14160-6075 based prototype) shows the expected strong dependence on overvoltage and on the trigger threshold. The rate $R_{\rm trig}$ is determined by using the rate for the number of fired pixels that correspond to 0.5\,MeV$_{ee}$ deposited energy, at the overvoltage used. 

\begin{table}[t]
\begin{tabular}{lD{.}{.}{1}rrr}
 & \multicolumn{1}{c}{$U_{\rm OP}$}    
 & $ R_\text{dark}^\text{obs} $ 
 & \multicolumn{1}{c}{ $R_{\rm trig}^\text{obs}$ }
 & \multicolumn{1}{c}{ $R_\text{dp}^\text{calc} $}  \\ 
& {[}V{]}
& {[} $10^{6} \ \text{s}^{-1}$ {]}
& \multicolumn{1}{c}{{[} $\text{s}^{-1}$ {]}}
&\multicolumn{1}{c}{ {[}$\text{s}^{-1}$ {]}}
  \\ \hline

{\tt onsemi}   & 27.0 & 16 & 20    & $<10$ \\
J Series 60035                          & 28.0 & 13 & 10    & $<10$ \\
                          & 29.0* & 15 & 70    & 10    \\
                          & 30.0 & 19 & 600    & 900    \\
                          & 31.0 & 26 & 4 400    & 47 000   \\ \hline
                          
Hamamatsu      & 41.0 & 19 & 2 500 & 14 000 \\
S14160-6050                        & 41.5* & 17 & 1 200 & 3 600 \\
                          & 42.0 & 17 & 700 & 1 000  \\
                          & 43.0 & 20 & 1 500 & 5 400  \\ \hline

Hamamatsu      & 40.0 & 17 & 1 100 & 2 900  \\
S14160-6075                          & 41.0* & 20 & 60 & 10  \\
                          & 41.5 & 21 & 200 & 90  \\
                          & 42.0 & 27 & 1 000 & 2 500  \\
                          & 43.0 & 38 & 4 400 & 47 000 \\ \hline
                          
Hamamatsu      & 56.0 & 10 & 150 & 50  \\
S13360-6075                          & 57.0 & 14 & 1 600 & 6 300  \\ \hline
\end{tabular}
\caption{Dark rates for the different SiPM models. The values are given for an array of four SiPMs at room temperature ($20^\circ\text{C}$). The overvoltages marked with an asterisk correspond to the ones shown in \autoref{fig:sat_SiPM}.}
\label{tab:dark_rates}
\end{table}

The results for the dark count rates of the SiPMs studied here are given in \autoref{tab:dark_rates}: For each overvoltage studied, the measured rate per prototype $R_{\rm trig}^{\rm obs}$ and the calculated dark rate $R_{\rm dp}^{\rm calc}$ using (\autoref{eq:darks}) and $R_{\rm trig}^{\rm obs}$ are shown. 

For the cases of the {\tt onsemi} J Series 60035 and Hamamatsu S14160-6075 SiPMs, a calculated dark rate of $R_\text{dp}^\text{calc} $ = 10 s$^{-1}$ per double plane is found. For a full NeuLAND with 30 double planes, this would mean a dark rate of 300 s$^{-1}$, lower than the expected cosmic rate of 1500 s$^{-1}$ for the full NeuLAND \cite{Boretzky21-NIMA}. The Hamamatsu S14160-6050 based prototype, however, shows a much higher dark rate.

For information, also the total dark count rate $R_{\rm dark}^{\rm obs}$ with a trigger threshold corresponding to half the amplitude of one fired pixel is shown. This value can be compared to the datasheet values of $R_{\text{dark}}$ (\autoref{tab:sipm_types}). The present measured $R_{\rm dark}^{\rm obs}$ is higher than the datasheet value for the {\tt onsemi} SiPM and lower for the Hamamatsu SiPMs, probably reflecting variability in this parameter from device to device. 

No study of the temperature dependence of the dark count rate has been performed here. Previous work has shown that the dark count rate doubles for every 5-10\,K temperature increase \cite{Otte17-NIMA}.

\subsection{Cosmic-ray muon response studied with the NeuLAND electronic readout at GSI} \label{subsec:GSI}

\begin{figure}[t]
    \centering
    \includegraphics[width=0.45\textwidth]{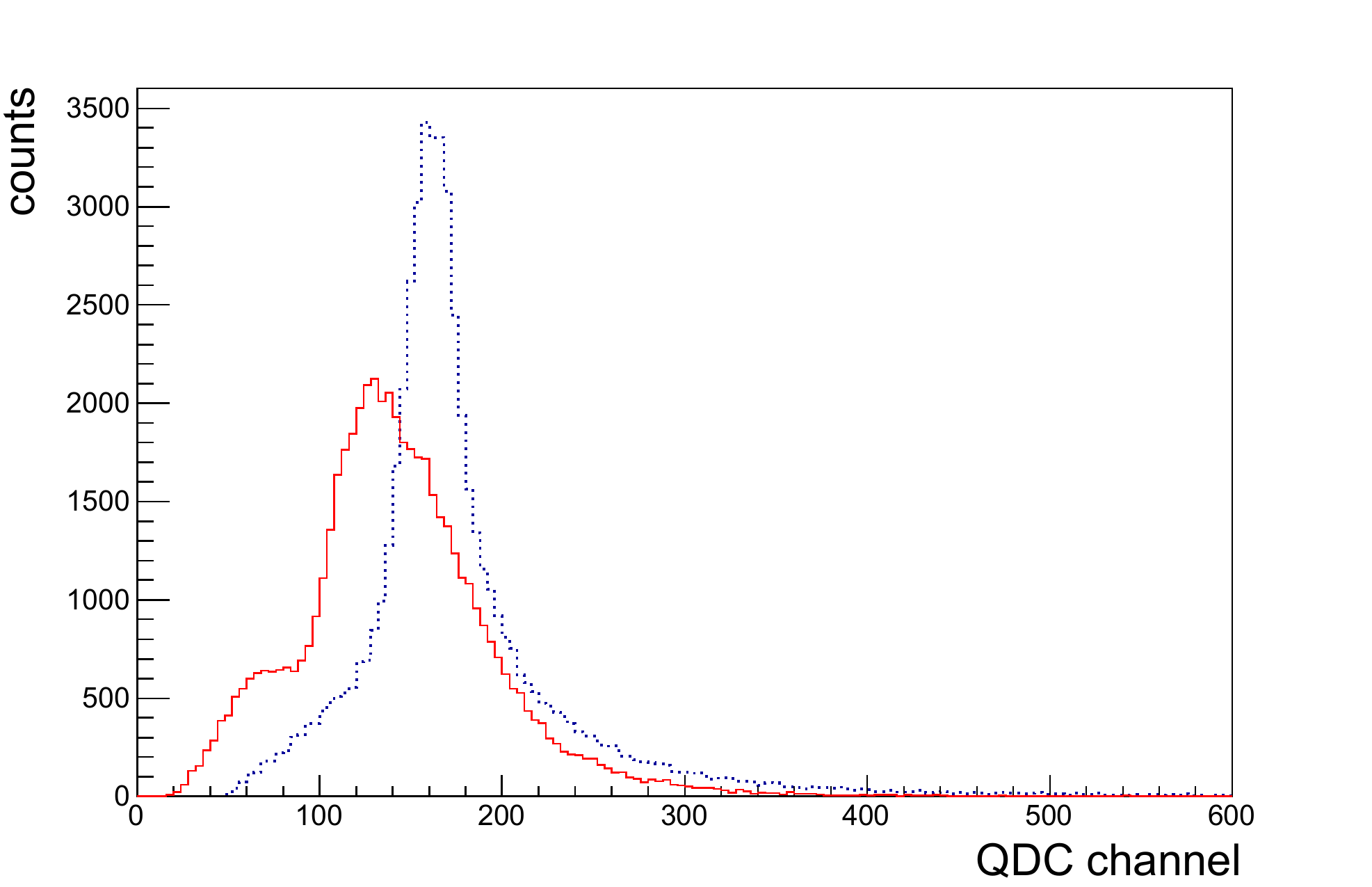}
    \caption{QDC spectrum of a NeuLAND bar equipped with a SiPM device (Hamamatsu S14160-6075, blue dotted line) and a standard Hamamatsu PMT (red line), recorded with the NeuLAND electronics. The trigger condition is set to a fourfold coincidence with a scintillator bar placed directly below selecting solely events generated by cosmic muons.}
    \label{fig:GSI_Test}
\end{figure}

In order to further evaluate the practicability of reading out the NeuLAND scintillation light by SiPMs, a test was performed at the NeuLAND test station of GSI, using the NeuLAND electronics (\autoref{subsec:ExpGSI}) and one Hamamatsu S14160-6075 based prototype. 

The observed charge spectrum (\autoref{fig:GSI_Test}, blue curve) for incident cosmic-ray muons generally resembles that of a standard PMT (\autoref{fig:GSI_Test}, red curve). There are two main differences. First, the SiPM-based prototype has about 20\% higher gain, a fact that can in principle be corrected using lower gain in the SiPM electronics. Second, even after correcting for the gain, the SiPM based prototype has a higher effective trigger threshold . This higher threshold is probably due to the fact that the total length of the pulses from the present prototype seems to be slightly too long, reducing the trigger efficiency especially for smaller signals. Again, this can in principle be corrected by a slight modification in the amplifier.

No measurement of the time resolutions with muons was attempted, as this would have required the use of a reference detector with essentially the same time resolution as the device under study. Instead, the time resolution was measured with a picosecond laser system, setting the amplitude to correspond to the main muon peak (\autoref{fig:GSI_Test}). For readout with the PMT, the resultant time resolution is $\sigma_t$ = 80\,ps. For readout with the SiPM, $\sigma_t$ = 83-100\,ps is found, depending on the operating voltage.

\section{Discussion}
\label{sec:Discussion}

In the present section, the various aspects needed for usage of a SiPM in NeuLAND that have been studied in the present work are discussed.

\begin{enumerate}
    \item The time resolution $\sigma_t$ is of fundamental importance for a time of flight detector such as NeuLAND, and the technical design report calls for $\sigma_t < 150$\,ps. In previous work, it had already been shown experimentally that this aim can be achieved with an array of four 6$\times$6\,mm$^2$ SiPMs, with an observed time resolution of $\sigma_t$ = 136\,ps \cite{Reinhardt16-NIMA}. Based on improvements in the prototype and also in the SiPMs studied, the present work exceeds this result. For most of the prototypes studied, there is a several volts wide plateau with $\sigma_t<120$\,ps  (\autoref{fig:sigma_vs_vop}). The one exception is the {\tt onsemi} (formerly SensL) C-series 60035 SiPM that had already been studied previously \cite{Reinhardt16-NIMA}. For this prototype, the best value is $\sigma_t$ = 142$\pm$5\,ps, consistent with the previous value of 136$\pm$2\,ps \cite{Reinhardt16-NIMA}.
    \item The detection efficiency for minimum ionizing particles was previously found to be $\geq$95\% for arrays of four 6$\times$6\,mm$^2$ SiPMs \cite{Reinhardt16-NIMA}, as required by NeuLAND. This value was again confirmed in the present work for all SiPMs studied here  (\autoref{fig:sigma_vs_vop}). 
    \item The dark rate has not been previously studied for SiPM arrays that will be used to detect high energy neutrons. The present results, when extrapolated to the realistic NeuLAND trigger conditions, show dark trigger rates ($R_\text{dp}^\text{calc}$) of ca. 100\,s$^{-1}$ for all prototypes studied at the optimum overvoltage (\autoref{tab:dark_rates}). There is one notable exception, the Hamamatsu S14160-6050 based prototype which displays an extrapolated dark trigger rate of 3600\,s$^{-1}$ for a fully equipped double plane, much higher but still tolerable for the NeuLAND DAQ system. The fact that these dark trigger rates are tolerable is owed to the relatively restrictive NeuLAND trigger condition of two photosensors simultaneously detecting above a $>0.5$\,MeV$_{ee}$ threshold.
    \item The linearity of the response of the SiPM + preamplifier setup was studied using the one to few electrons per bunch of ELBE. It was found that for a deposited energy of up to 300\,MeV per NeuLAND bar, the total signal deviates by $\leq$20\% from linearity, and the slope at this highest energies is reduced by $\leq$35\% for all cases (\autoref{fig:sat_SiPM}). This relatively modest deviation from linearity will result in only 10\% degradation of the energy resolution and is still compatible with a reliable energy calibration.
    \item Finally, the practicality of connecting a NeuLAND bar read out by the present SiPM-based prototype to the NeuLAND electronics was studied using cosmic-ray and laser-based tests. It was shown that also with the NeuLAND electronics the present prototype functions well within the requested performance envelope of NeuLAND photosensors.
\end{enumerate}

\section{Summary and outlook}
\label{sec:Summary}

Motivated by the question whether a large scintillator-based time-of-flight detector such as NeuLAND can be successfully instrumented with semiconductor-based photosensors, called silicon photomultipliers (SiPMs), a new prototype hosting four SiPMs was constructed. This prototype was designed to replace a 1'' classical photomultiplier tube for usage at one end of a NeuLAND plastic scintillator bar.

Extensive tests have been done with picosecond timing lasers, cosmic-ray muons and above all single or few-electron bunches of 35\,MeV electrons to study the efficiency, time resolution, dark trigger rate, linearity, and matching to the classical NeuLAND data acquisition system. All parameters tested were found to be within the performance envelope required for NeuLAND, for SiPMs operated at room temperature without active temperature stabilisation. 

Remaining questions include the practicability to manufacture a SiPM-based readout device on the scale required to (re)instrument NeuLAND with its 6,000 photosensors, and economic considerations. At current market prices, for the SiPMs studied here, the cost of the SiPM-based readout ranges from below to above the PMT option, depending on the prototype and manufacturer chosen. Also, for the specific case of NeuLAND, it has to be recognised that the SiPM operating voltages are much below the ones for the PMT option adopted, presenting advantages for safety and cabling.

It is hoped that this work, beyond its potential for NeuLAND, may also assist other groups pursuing similar detectors \cite[e.g.]{NOVO18-ARX}

\section*{Acknowledgements} \label{sec:acknowledge}
The authors are indebted to Andreas Hartmann (HZDR) for technical support and to Katja Römer (HZDR), Benjanin Lutz (HZDR, now PTB), Christof Schneider (HZDR) and Jörg Pawelke (Oncoray) for helpful discussions. -- Financial support by GSI (F\&E DD-ZUBE, 2020-2022), by BMBF (NupNET NEDENSAA 05P09CRFN5, 2012-2015), and by the European Union (ChETEC-INFRA, 101008324) is gratefully acknowledged. I.G. acknowledges support by the Croatian Science Foundation under project number 1257.

\section*{Dedication}
This work is dedicated to the memory of our late colleague Dr.-Ing. David Weinberger.


\begin{thebibliography}{10}
\expandafter\ifx\csname url\endcsname\relax
  \def\url#1{\texttt{#1}}\fi
\expandafter\ifx\csname urlprefix\endcsname\relax\def\urlprefix{URL }\fi
\expandafter\ifx\csname href\endcsname\relax
  \def\href#1#2{#2} \def\path#1{#1}\fi

\bibitem{Buzhan03-NIMA}
P.~Buzhan, B.~Dolgoshein, L.~Filatov, A.~Ilyin, V.~Kantzerov, V.~Kaplin,
  A.~Karakash, F.~Kayumov, S.~Klemin, E.~Popova, S.~Smirnov,
  \href{https://www.sciencedirect.com/science/article/pii/S0168900203007496}{Silicon
  photomultiplier and its possible applications}, Nucl.~Inst.~Meth.~A 504~(1)
  (2003) 48--52.
\newblock \href
  {http://dx.doi.org/https://doi.org/10.1016/S0168-9002(03)00749-6}
  {\path{doi:https://doi.org/10.1016/S0168-9002(03)00749-6}}.
\newline\urlprefix\url{https://www.sciencedirect.com/science/article/pii/S0168900203007496}

\bibitem{Nakamura16-NIMB}
T.~{Nakamura}, Y.~{Kondo},
  \href{https://www.sciencedirect.com/science/article/pii/S0168583X1600029X}{Large
  acceptance spectrometers for invariant mass spectroscopy of exotic nuclei and
  future developments}, Nucl.~Inst.~Meth.~B 376 (2016) 156--161.
\newblock \href {http://dx.doi.org/https://doi.org/10.1016/j.nimb.2016.01.003}
  {\path{doi:https://doi.org/10.1016/j.nimb.2016.01.003}}.
\newline\urlprefix\url{https://www.sciencedirect.com/science/article/pii/S0168583X1600029X}

\bibitem{Wang19-NIMA}
C.~Wang, X.~Li, R.~Han, Z.~Li, H.~Hua, S.~Zhang, D.~Jiang, Y.~Ye, J.~Li, Z.~Li,
  J.~Wang, C.~Xu, Y.~Sun, H.~Wu, C.~Niu, C.~Li, Z.~Chen, D.~Luo, C.~He,
  W.~Jiang, P.~Li, H.~Zang, J.~Feng, S.~Chen, Q.~Liu, X.~Chen, H.~Xu, Z.~Hu,
  Y.~Yang, P.~Ma, J.~Ma, S.~Jin, Z.~Bai, M.~Huang, Y.~Zhou, W.~Ma, Y.~Li,
  X.~Zhou, Y.~Zhang, G.~Xiao, W.~Zhan,
  \href{https://www.sciencedirect.com/science/article/pii/S0168900219308265}{A
  β-delayed neutron detection system working with the continuous beam mode},
  Nucl.~Inst.~Meth.~A 940 (2019) 83--87.
\newblock \href {http://dx.doi.org/https://doi.org/10.1016/j.nima.2019.06.008}
  {\path{doi:https://doi.org/10.1016/j.nima.2019.06.008}}.
\newline\urlprefix\url{https://www.sciencedirect.com/science/article/pii/S0168900219308265}

\bibitem{Reinhardt16-NIMA}
T.~Reinhardt, S.~Gohl, S.~Reinicke, D.~Bemmerer, T.~Cowan, K.~Heidel,
  M.~Röder, D.~Stach, A.~Wagner, D.~Weinberger, K.~Zuber, Silicon
  photomultiplier readout of a monolithic 270$\times$5$\times$5 cm$^3$ plastic
  scintillator bar for time of flight applications, Nucl.~Inst.~Meth.~A 816.
\newblock \href {http://dx.doi.org/10.1016/j.nima.2016.01.054}
  {\path{doi:10.1016/j.nima.2016.01.054}}.

\bibitem{Wojtowicz20-Polo}
P.~Wojtowicz, K.~Miernik, A.~Korgul, M.~Piersa-Siłkowska, M.~Siłkowski,
  T.~Rogiński,
  \href{https://www.actaphys.uj.edu.pl/fulltext?series=Reg&vol=51&page=725}{Warsaw
  time-of-flight neutron detector}, Acta Physica Polonica B 51 (2020) 725.
\newblock \href {http://dx.doi.org/10.5506/APhysPolB.51.725}
  {\path{doi:10.5506/APhysPolB.51.725}}.
\newline\urlprefix\url{https://www.actaphys.uj.edu.pl/fulltext?series=Reg&vol=51&page=725}

\bibitem{Simon19-NIMA}
F.~Simon,
  \href{https://www.sciencedirect.com/science/article/pii/S0168900218316176}{Silicon
  photomultipliers in particle and nuclear physics}, Nucl.~Inst.~Meth.~A 926
  (2019) 85--100, silicon Photomultipliers: Technology, Characterisation and
  Applications.
\newblock \href {http://dx.doi.org/https://doi.org/10.1016/j.nima.2018.11.042}
  {\path{doi:https://doi.org/10.1016/j.nima.2018.11.042}}.
\newline\urlprefix\url{https://www.sciencedirect.com/science/article/pii/S0168900218316176}

\bibitem{Boehm16-JOI}
M.~Böhm, A.~Lehmann, S.~Motz, F.~Uhlig,
  \href{https://doi.org/10.1088/1748-0221/11/05/c05018}{Fast {SiPM} readout of
  the {PANDA} {TOF} detector}, Journal of Instrumentation 11~(05) (2016)
  C05018--C05018.
\newblock \href {http://dx.doi.org/10.1088/1748-0221/11/05/c05018}
  {\path{doi:10.1088/1748-0221/11/05/c05018}}.
\newline\urlprefix\url{https://doi.org/10.1088/1748-0221/11/05/c05018}

\bibitem{DeGuio19-JOP}
F.~D. Guio, {on behalf of the CMS Collaboration},
  \href{https://doi.org/10.1088/1742-6596/1162/1/012009}{First results from the
  {CMS} {SiPM}-based hadronic endcap calorimeter}, Journal of Physics:
  Conference Series 1162 (2019) 012009.
\newblock \href {http://dx.doi.org/10.1088/1742-6596/1162/1/012009}
  {\path{doi:10.1088/1742-6596/1162/1/012009}}.
\newline\urlprefix\url{https://doi.org/10.1088/1742-6596/1162/1/012009}

\bibitem{Strobbe17-JOI}
N.~Strobbe, \href{https://doi.org/10.1088/1748-0221/12/01/c01080}{The upgrade
  of the {CMS} hadron calorimeter with silicon photomultipliers}, Journal of
  Instrumentation 12~(01) (2017) C01080--C01080.
\newblock \href {http://dx.doi.org/10.1088/1748-0221/12/01/c01080}
  {\path{doi:10.1088/1748-0221/12/01/c01080}}.
\newline\urlprefix\url{https://doi.org/10.1088/1748-0221/12/01/c01080}

\bibitem{Boretzky21-NIMA}
K.~{Boretzky}, I.~{Ga{\v{s}}pari{\'c}}, M.~{Heil}, J.~{Mayer}, A.~{Heinz},
  C.~{Caesar}, D.~{Kresan}, H.~{Simon}, H.~T. {T{\"o}rnqvist}, D.~{K{\"o}rper},
  G.~{Alkhazov}, L.~{Atar}, T.~{Aumann}, D.~{Bemmerer}, S.~V. {Bondarev}, L.~T.
  {Bott}, S.~{Chakraborty}, M.~I. {Cherciu}, L.~V. {Chulkov}, M.~{Ciobanu},
  U.~{Datta}, E.~{De Filippo}, C.~A. {Douma}, J.~{Dreyer}, Z.~{Elekes},
  J.~{Enders}, D.~{Galaviz}, E.~{Geraci}, B.~{Gnoffo}, K.~{G{\"o}bel}, V.~L.
  {Golovtsov}, D.~{Gonzalez Diaz}, N.~{Gruzinsky}, T.~{Heftrich}, H.~{Heggen},
  J.~{Hehner}, T.~{Hensel}, E.~{Hoemann}, M.~{Holl}, A.~{Horvat},
  {\'A}.~{Horv{\'a}th}, G.~{Ickert}, D.~{Jelavi{\'c} Malenica}, H.~T.
  {Johansson}, B.~{Jonson}, J.~{Kahlbow}, N.~{Kalantar-Nayestanaki},
  A.~{Keli{\'c}-Heil}, M.~{Kempe}, K.~{Koch}, N.~G. {Kozlenko}, A.~G.
  {Krivshich}, N.~{Kurz}, V.~{Kuznetsov}, C.~{Langer}, Y.~{Leifels},
  I.~{Lihtar}, B.~{L{\"o}her}, J.~{Machado}, N.~S. {Martorana}, K.~{Miki},
  T.~{Nilsson}, E.~M. {Orischin}, E.~V. {Pagano}, S.~{Pirrone}, G.~{Politi},
  P.~M. {Potlog}, A.~{Rahaman}, R.~{Reifarth}, C.~{Rigollet}, M.~{R{\"o}der},
  D.~M. {Rossi}, P.~{Russotto}, D.~{Savran}, H.~{Scheit}, F.~{Schindler},
  D.~{Stach}, E.~{Stan}, J.~{Stomvall Gill}, P.~{Teubig}, M.~{Trimarchi},
  L.~{Uvarov}, M.~{Volknandt}, S.~{Volkov}, A.~{Wagner}, V.~{Wagner},
  S.~{Wranne}, D.~{Yakorev}, L.~{Zanetti}, A.~{Zilges}, K.~{Zuber}, {R$^{3}$B
  collaboration}, {NeuLAND: The high-resolution neutron time-of-flight
  spectrometer for R$^{3}$B at FAIR}, Nuclear Instruments and Methods in
  Physics Research A 1014 (2021) 165701.
\newblock \href {http://dx.doi.org/10.1016/j.nima.2021.165701}
  {\path{doi:10.1016/j.nima.2021.165701}}.

\bibitem{Mayer21-NIMA}
J.~Mayer, K.~Boretzky, C.~Douma, E.~Hoemann, A.~Zilges,
  \href{https://repository.gsi.de/record/240860}{{C}lassical and machine
  learning methods for event reconstruction in {N}eu{LAND}},
  Nucl.~Inst.~Meth.~A 1013.
\newblock \href {http://dx.doi.org/10.1016/j.nima.2021.165666}
  {\path{doi:10.1016/j.nima.2021.165666}}.
\newline\urlprefix\url{https://repository.gsi.de/record/240860}

\bibitem{Douma21-NIMA}
C.~Douma, E.~Hoemann, N.~Kalantar-Nayestanaki, J.~Mayer,
  \href{https://www.sciencedirect.com/science/article/pii/S0168900220313486}{Development
  of a deep neural network for the data analysis of the neuland neutron
  detector}, Nuclear Instruments and Methods in Physics Research Section A:
  Accelerators, Spectrometers, Detectors and Associated Equipment 990 (2021)
  164951.
\newblock \href {http://dx.doi.org/https://doi.org/10.1016/j.nima.2020.164951}
  {\path{doi:https://doi.org/10.1016/j.nima.2020.164951}}.
\newline\urlprefix\url{https://www.sciencedirect.com/science/article/pii/S0168900220313486}

\bibitem{Roeder16-PRC}
M.~{R{\"o}der}, T.~{Adachi}, Y.~{Aksyutina}, J.~{Alcantara}, S.~{Altstadt},
  H.~{Alvarez-Pol}, N.~{Ashwood}, L.~{Atar}, T.~{Aumann}, V.~{Avdeichikov},
  M.~{Barr}, S.~{Beceiro}, D.~{Bemmerer}, J.~{Benlliure}, C.~{Bertulani},
  K.~{Boretzky}, M.~J.~G. {Borge}, G.~{Burgunder}, M.~{Caama{\~n}o},
  C.~{Caesar}, E.~{Casarejos}, W.~{Catford}, J.~{Cederk{\"a}ll},
  S.~{Chakraborty}, M.~{Chartier}, L.~{Chulkov}, D.~{Cortina-Gil}, R.~{Crespo},
  U.~{Datta Pramanik}, P.~{Diaz-Fernandez}, I.~{Dillmann}, Z.~{Elekes},
  J.~{Enders}, O.~{Ershova}, A.~{Estrade}, F.~{Farinon}, L.~M. {Fraile},
  M.~{Freer}, M.~{Freudenberger}, H.~{Fynbo}, D.~{Galaviz}, H.~{Geissel},
  R.~{Gernh{\"a}user}, K.~{G{\"o}bel}, P.~{Golubev}, D.~{Gonzalez Diaz},
  J.~{Hagdahl}, T.~{Heftrich}, M.~{Heil}, M.~{Heine}, A.~{Heinz},
  A.~{Henriques}, M.~{Holl}, G.~{Ickert}, A.~{Ignatov}, B.~{Jakobsson},
  H.~{Johansson}, B.~{Jonson}, N.~{Kalantar-Nayestanaki}, R.~{Kanungo},
  A.~{Kelic-Heil}, R.~{Kn{\"o}bel}, T.~{Kr{\"o}ll}, R.~{Kr{\"u}cken},
  J.~{Kurcewicz}, N.~{Kurz}, M.~{Labiche}, C.~{Langer}, T.~{Le Bleis},
  R.~{Lemmon}, O.~{Lepyoshkina}, S.~{Lindberg}, J.~{Machado}, J.~{Marganiec},
  M.~{Mostazo Caro}, A.~{Movsesyan}, M.~A. {Najafi}, T.~{Nilsson},
  C.~{Nociforo}, V.~{Panin}, S.~{Paschalis}, A.~{Perea}, M.~{Petri},
  S.~{Pietri}, R.~{Plag}, A.~{Prochazka}, M.~A. {Rahaman}, G.~{Rastrepina},
  R.~{Reifarth}, G.~{Ribeiro}, M.~V. {Ricciardi}, C.~{Rigollet}, K.~{Riisager},
  D.~{Rossi}, J.~{Sanchez del Rio Saez}, D.~{Savran}, H.~{Scheit}, H.~{Simon},
  O.~{Sorlin}, V.~{Stoica}, B.~{Streicher}, J.~{Taylor}, O.~{Tengblad},
  S.~{Terashima}, R.~{Thies}, Y.~{Togano}, E.~{Uberseder}, J.~{Van de Walle},
  P.~{Velho}, V.~{Volkov}, A.~{Wagner}, F.~{Wamers}, H.~{Weick}, M.~{Weigand},
  C.~{Wheldon}, G.~{Wilson}, C.~{Wimmer}, J.~S. {Winfield}, P.~{Woods},
  D.~{Yakorev}, M.~{Zhukov}, A.~{Zilges}, K.~{Zuber}, {R$^3$B Collaboration},
  {Coulomb dissociation of $^{20,21}$N}, Phys.~Rev.~C 93~(6) (2016) 065807.
\newblock \href {http://dx.doi.org/10.1103/PhysRevC.93.065807}
  {\path{doi:10.1103/PhysRevC.93.065807}}.

\bibitem{Heine17-PRC}
M.~{Heine}, S.~{Typel}, M.~R. {Wu}, T.~{Adachi}, Y.~{Aksyutina},
  J.~{Alcantara}, S.~{Altstadt}, H.~{Alvarez-Pol}, N.~{Ashwood}, L.~{Atar},
  T.~{Aumann}, V.~{Avdeichikov}, M.~{Barr}, S.~{Beceiro-Novo}, D.~{Bemmerer},
  J.~{Benlliure}, C.~A. {Bertulani}, K.~{Boretzky}, M.~J.~G. {Borge},
  G.~{Burgunder}, M.~{Caamano}, C.~{Caesar}, E.~{Casarejos}, W.~{Catford},
  J.~{Cederk{\"a}ll}, S.~{Chakraborty}, M.~{Chartier}, L.~V. {Chulkov},
  D.~{Cortina-Gil}, R.~{Crespo}, U.~{Datta Pramanik}, P.~{Diaz Fernandez},
  I.~{Dillmann}, Z.~{Elekes}, J.~{Enders}, O.~{Ershova}, A.~{Estrade},
  F.~{Farinon}, L.~M. {Fraile}, M.~{Freer}, M.~{Freudenberger}, H.~O.~U.
  {Fynbo}, D.~{Galaviz}, H.~{Geissel}, R.~{Gernh{\"a}user}, K.~{G{\"o}bel},
  P.~{Golubev}, D.~{Gonzalez Diaz}, J.~{Hagdahl}, T.~{Heftrich}, M.~{Heil},
  A.~{Heinz}, A.~{Henriques}, M.~{Holl}, G.~{Ickert}, A.~{Ignatov},
  B.~{Jakobsson}, H.~T. {Johansson}, B.~{Jonson}, N.~{Kalantar-Nayestanaki},
  R.~{Kanungo}, A.~{Kelic-Heil}, R.~{Kn{\"o}bel}, T.~{Kr{\"o}ll},
  R.~{Kr{\"u}cken}, J.~{Kurcewicz}, N.~{Kurz}, M.~{Labiche}, C.~{Langer},
  T.~{Le Bleis}, R.~{Lemmon}, O.~{Lepyoshkina}, S.~{Lindberg}, J.~{Machado},
  J.~{Marganiec}, G.~{Mart{\'\i}nez-Pinedo}, V.~{Maroussov}, M.~{Mostazo},
  A.~{Movsesyan}, A.~{Najafi}, T.~{Neff}, T.~{Nilsson}, C.~{Nociforo},
  V.~{Panin}, S.~{Paschalis}, A.~{Perea}, M.~{Petri}, S.~{Pietri}, R.~{Plag},
  A.~{Prochazka}, A.~{Rahaman}, G.~{Rastrepina}, R.~{Reifarth}, G.~{Ribeiro},
  M.~V. {Ricciardi}, C.~{Rigollet}, K.~{Riisager}, M.~{R{\"o}der}, D.~{Rossi},
  J.~{Sanchez del Rio}, D.~{Savran}, H.~{Scheit}, H.~{Simon}, O.~{Sorlin},
  V.~{Stoica}, B.~{Streicher}, J.~T. {Taylor}, O.~{Tengblad}, S.~{Terashima},
  R.~{Thies}, Y.~{Togano}, E.~{Uberseder}, J.~{Van de Walle}, P.~{Velho},
  V.~{Volkov}, A.~{Wagner}, F.~{Wamers}, H.~{Weick}, M.~{Weigand},
  C.~{Wheldon}, G.~{Wilson}, C.~{Wimmer}, J.~S. {Winfield}, P.~{Woods},
  D.~{Yakorev}, M.~V. {Zhukov}, A.~{Zilges}, K.~{Zuber}, {R3B Collaboration},
  {Determination of the neutron-capture rate of $^{17}$C for r -process
  nucleosynthesis}, \prc 95~(1) (2017) 014613.
\newblock \href {http://arxiv.org/abs/1604.05832} {\path{arXiv:1604.05832}},
  \href {http://dx.doi.org/10.1103/PhysRevC.95.014613}
  {\path{doi:10.1103/PhysRevC.95.014613}}.

\bibitem{Duer22-Nature}
M.~Duer, T.~Aumann, R.~Gernh{\"a}user, V.~Panin, S.~Paschalis, D.~M. Rossi,
  N.~L. Achouri, D.~Ahn, H.~Baba, C.~A. Bertulani, M.~B{\"o}hmer, K.~Boretzky,
  C.~Caesar, N.~Chiga, A.~Corsi, D.~Cortina-Gil, C.~A. Douma, F.~Dufter,
  Z.~Elekes, J.~Feng, B.~Fern{\'a}ndez-Dom{\'\i}nguez, U.~Forsberg, N.~Fukuda,
  I.~Gasparic, Z.~Ge, J.~M. Gheller, J.~Gibelin, A.~Gillibert, K.~I. Hahn,
  Z.~Hal{\'a}sz, M.~N. Harakeh, A.~Hirayama, M.~Holl, N.~Inabe, T.~Isobe,
  J.~Kahlbow, N.~Kalantar-Nayestanaki, D.~Kim, S.~Kim, T.~Kobayashi, Y.~Kondo,
  D.~K{\"o}rper, P.~Koseoglou, Y.~Kubota, I.~Kuti, P.~J. Li, C.~Lehr,
  S.~Lindberg, Y.~Liu, F.~M. Marqu{\'e}s, S.~Masuoka, M.~Matsumoto, J.~Mayer,
  K.~Miki, B.~Monteagudo, T.~Nakamura, T.~Nilsson, A.~Obertelli, N.~A. Orr,
  H.~Otsu, S.~Y. Park, M.~Parlog, P.~M. Potlog, S.~Reichert, A.~Revel, A.~T.
  Saito, M.~Sasano, H.~Scheit, F.~Schindler, S.~Shimoura, H.~Simon, L.~Stuhl,
  H.~Suzuki, D.~Symochko, H.~Takeda, J.~Tanaka, Y.~Togano, T.~Tomai, H.~T.
  T{\"o}rnqvist, J.~Tscheuschner, T.~Uesaka, V.~Wagner, H.~Yamada, B.~Yang,
  L.~Yang, Z.~H. Yang, M.~Yasuda, K.~Yoneda, L.~Zanetti, J.~Zenihiro, M.~V.
  Zhukov, \href{https://doi.org/10.1038/s41586-022-04827-6}{Observation of a
  correlated free four-neutron system}, Nature 606~(7915) (2022) 678--682.
\newblock \href {http://dx.doi.org/10.1038/s41586-022-04827-6}
  {\path{doi:10.1038/s41586-022-04827-6}}.
\newline\urlprefix\url{https://doi.org/10.1038/s41586-022-04827-6}

\bibitem{Yakorev11-NIMA}
D.~Yakorev, T.~Aumann, D.~Bemmerer, K.~Boretzky, C.~Caesar, M.~Ciobanu,
  T.~Cowan, Z.~Elekes, M.~Elvers, D.~G. Diaz, R.~Hannaske, J.~Hehner, M.~Heil,
  M.~Kempe, V.~Maroussov, O.~Nusair, H.~Simon, M.~Sobiella, D.~Stach,
  A.~Wagner, A.~Zilges, {Prototyping and tests for an MRPC-based time-of-flight
  detector for 1 GeV neutrons}, Nucl.~Inst.~Meth.~A 654~(1) (2011) 79 -- 87.
\newblock \href {http://dx.doi.org/10.1016/j.nima.2011.05.031}
  {\path{doi:10.1016/j.nima.2011.05.031}}.

\bibitem{Roeder12-JINST}
M.~{R{\"o}der}, T.~{Aumann}, D.~{Bemmerer}, K.~{Boretzky}, C.~{Caesar}, T.~E.
  {Cowan}, J.~{Hehner}, M.~{Heil}, Z.~{Elekes}, M.~{Kempe}, V.~{Maroussov},
  T.~P. {Reinhardt}, H.~{Simon}, M.~{Sobiella}, D.~{Stach}, A.~{Wagner},
  J.~{W{\"u}stenfeld}, D.~{Yakorev}, A.~{Zilges}, K.~{Zuber}, {Prototyping a
  2m$\times$0.5m MRPC-based neutron TOF-wall with steel converter plates},
  J.~Inst. 7 (2012) 1030P.
\newblock \href {http://dx.doi.org/10.1088/1748-0221/7/11/P11030}
  {\path{doi:10.1088/1748-0221/7/11/P11030}}.

\bibitem{Elekes13-NIMA}
Z.~{Elekes}, T.~{Aumann}, D.~{Bemmerer}, K.~{Boretzky}, C.~{Caesar}, T.~C.
  {Cowan}, J.~{Hehner}, M.~{Heil}, M.~{Kempe}, D.~{Rossi}, M.~{R{\"o}der},
  H.~{Simon}, M.~{Sobiella}, D.~{Stach}, T.~{Reinhardt}, A.~{Wagner},
  D.~{Yakorev}, A.~{Zilges}, K.~{Zuber}, {Simulation and prototyping of 2 m
  long resistive plate chambers for detection of fast neutrons and
  multi-neutron event identification}, Nucl.~Inst.~Meth.~A 701 (2013) 86--92.
\newblock \href {http://dx.doi.org/10.1016/j.nima.2012.11.010}
  {\path{doi:10.1016/j.nima.2012.11.010}}.

\bibitem{NeuLAND11-TDR}
{R$^3$B collaboration},
  \href{{https://edms.cern.ch/ui/file/1865739/2/TDR_R3B_NeuLAND_public.pdf}}{{NeuLAND
  Technical Design Report}} (2011).
\newline\urlprefix\url{{https://edms.cern.ch/ui/file/1865739/2/TDR_R3B_NeuLAND_public.pdf}}

\bibitem{Douma19-NIMA}
C.~A. {Douma}, K.~{Boretzky}, I.~{Ga{\v{s}}pari{\'c}},
  N.~{Kalantar-Nayestanaki}, D.~{Kresan}, J.~{Mayer}, C.~{Rigollet}, {R3B
  Collaboration}, {Investigation of background reduction techniques for the
  NeuLAND neutron detector}, Nuclear Instruments and Methods in Physics
  Research A 930 (2019) 203--209.
\newblock \href {http://dx.doi.org/10.1016/j.nima.2019.03.068}
  {\path{doi:10.1016/j.nima.2019.03.068}}.

\bibitem{Geant416-NIMA}
J.~{Allison et al.},
  \href{https://www.sciencedirect.com/science/article/pii/S0168900216306957}{Recent
  developments in geant4}, Nucl.~Inst.~Meth.~A 835 (2016) 186--225.
\newblock \href {http://dx.doi.org/https://doi.org/10.1016/j.nima.2016.06.125}
  {\path{doi:https://doi.org/10.1016/j.nima.2016.06.125}}.
\newline\urlprefix\url{https://www.sciencedirect.com/science/article/pii/S0168900216306957}

\bibitem{vanDerLaan10-PMB}
D.~J. {van der Laan}, D.~R. {Schaart}, M.~C. {Maas}, F.~J. {Beekman},
  P.~{Bruyndonckx}, C.~W.~E. {van Eijk}, {Optical simulation of monolithic
  scintillator detectors using GATE/GEANT4}, Physics in Medicine and Biology
  55~(6) (2010) 1659--1675.
\newblock \href {http://dx.doi.org/10.1088/0031-9155/55/6/009}
  {\path{doi:10.1088/0031-9155/55/6/009}}.

\bibitem{Hartwig14-NIMA}
Z.~S. {Hartwig}, P.~{Gumplinger}, {Simulating response functions and pulse
  shape discrimination for organic scintillation detectors with Geant4},
  Nucl.~Inst.~Meth.~A 737 (2014) 155--162.
\newblock \href {http://dx.doi.org/10.1016/j.nima.2013.11.027}
  {\path{doi:10.1016/j.nima.2013.11.027}}.

\bibitem{Karsch12-MedPhy}
L.~Karsch, E.~Beyreuther, T.~Burris-Mog, S.~Kraft, C.~Richter, K.~Zeil,
  J.~Pawelke,
  \href{https://aapm.onlinelibrary.wiley.com/doi/abs/10.1118/1.3700400}{Dose
  rate dependence for different dosimeters and detectors: Tld, osl, ebt films,
  and diamond detectors}, Medical Physics 39~(5) (2012) 2447--2455.
\newblock \href
  {http://arxiv.org/abs/https://aapm.onlinelibrary.wiley.com/doi/pdf/10.1118/1.3700400}
  {\path{arXiv:https://aapm.onlinelibrary.wiley.com/doi/pdf/10.1118/1.3700400}},
  \href {http://dx.doi.org/https://doi.org/10.1118/1.3700400}
  {\path{doi:https://doi.org/10.1118/1.3700400}}.
\newline\urlprefix\url{https://aapm.onlinelibrary.wiley.com/doi/abs/10.1118/1.3700400}

\bibitem{Kroll13-MedPhy}
F.~Kroll, J.~Pawelke, L.~Karsch,
  \href{https://aapm.onlinelibrary.wiley.com/doi/abs/10.1118/1.4813898}{Preliminary
  investigations on the determination of three-dimensional dose distributions
  using scintillator blocks and optical tomography}, Medical Physics 40~(8)
  (2013) 082104.
\newblock \href
  {http://arxiv.org/abs/https://aapm.onlinelibrary.wiley.com/doi/pdf/10.1118/1.4813898}
  {\path{arXiv:https://aapm.onlinelibrary.wiley.com/doi/pdf/10.1118/1.4813898}},
  \href {http://dx.doi.org/https://doi.org/10.1118/1.4813898}
  {\path{doi:https://doi.org/10.1118/1.4813898}}.
\newline\urlprefix\url{https://aapm.onlinelibrary.wiley.com/doi/abs/10.1118/1.4813898}

\bibitem{LASOGARCIA16-NIMA}
A.~{Laso Garcia}, R.~Kotte, L.~Naumann, D.~Stach, C.~Wendisch, J.~Wüstenfeld,
  B.~Kämpfer,
  \href{https://www.sciencedirect.com/science/article/pii/S0168900216001868}{High-rate
  timing resistive plate chambers with ceramic electrodes}, Nucl.~Inst.~Meth.~A
  818 (2016) 45--50.
\newblock \href {http://dx.doi.org/https://doi.org/10.1016/j.nima.2016.02.024}
  {\path{doi:https://doi.org/10.1016/j.nima.2016.02.024}}.
\newline\urlprefix\url{https://www.sciencedirect.com/science/article/pii/S0168900216001868}

\bibitem{Schwengner05-NIMA}
R.~Schwengner, R.~Beyer, F.~D{\"o}nau, E.~Grosse, A.~Hartmann, A.~Junghans,
  S.~Mallion, G.~Rusev, K.~Schilling, W.~Schulze, A.~Wagner, The
  photon-scattering facility at the superconducting electron accelerator elbe,
  Nucl.~Inst.~Meth.~A 555~(1-2) (2005) 211 -- 219.
\newblock \href {http://dx.doi.org/DOI: 10.1016/j.nima.2005.09.024}
  {\path{doi:DOI: 10.1016/j.nima.2005.09.024}}.

\bibitem{Beyer13-NIMA}
R.~{Beyer}, E.~{Birgersson}, Z.~{Elekes}, A.~{Ferrari}, E.~{Grosse},
  R.~{Hannaske}, A.~R. {Junghans}, T.~{K{\"o}gler}, R.~{Massarczyk},
  A.~{Mati{\'c}}, R.~{Nolte}, R.~{Schwengner}, A.~{Wagner}, {Characterization
  of the neutron beam at nELBE}, Nucl.~Inst.~Meth.~A 723 (2013) 151--162.
\newblock \href {http://dx.doi.org/10.1016/j.nima.2013.05.010}
  {\path{doi:10.1016/j.nima.2013.05.010}}.

\bibitem{Naumann11-NIMA}
L.~Naumann, R.~Kotte, D.~Stach, J.~W\"ustenfeld, {Ceramics high rate timing
  RPC}, Nucl.~Inst.~Meth.~A 628~(1) (2011) 138--141.

\bibitem{2022NIMPA103466745S}
R.~{Sekiya}, V.~{Drozd}, Y.~K. {Tanaka}, K.~{Itahashi}, H.~{Fujioka}, S.~Y.
  {Matsumoto}, T.~R. {Saito}, K.~{Suzuki}, {Time resolution and high-counting
  rate performance of plastic scintillation counter with multiple MPPC
  readout}, Nuclear Instruments and Methods in Physics Research A 1034 (2022)
  166745.
\newblock \href {http://dx.doi.org/10.1016/j.nima.2022.166745}
  {\path{doi:10.1016/j.nima.2022.166745}}.

\bibitem{2022JInst..17P1016K}
A.~{Korzenev}, F.~{Barao}, S.~{Bordoni}, D.~{Breton}, F.~{Cadoux}, Y.~{Favre},
  M.~{Khabibullin}, A.~{Khotyantsev}, Y.~{Kudenko}, T.~{Lux}, J.~{Maalmi},
  P.~{Mermod}, O.~{Mineev}, F.~{Sanchez}, {A 4{\ensuremath{\pi}} time-of-flight
  detector for the ND280/T2K upgrade}, Journal of Instrumentation 17~(1) (2022)
  P01016.
\newblock \href {http://arxiv.org/abs/2109.03078} {\path{arXiv:2109.03078}},
  \href {http://dx.doi.org/10.1088/1748-0221/17/01/P01016}
  {\path{doi:10.1088/1748-0221/17/01/P01016}}.

\bibitem{Gruber14-NIMA}
L.~{Gruber}, S.~E. {Brunner}, J.~{Marton}, K.~{Suzuki}, {Over saturation
  behavior of SiPMs at high photon exposure}, Nucl.~Inst.~Meth.~A 737 (2014)
  11--18.
\newblock \href {http://dx.doi.org/10.1016/j.nima.2013.11.013}
  {\path{doi:10.1016/j.nima.2013.11.013}}.

\bibitem{Weitzel19-NIMA}
Q.~{Weitzel}, et~al.,
  \href{https://www.sciencedirect.com/science/article/pii/S0168900218313883}{Measurement
  of the response of silicon photomultipliers from single photon detection to
  saturation}, Nucl.~Inst.~Meth.~A 936 (2019) 558--560.
\newblock \href {http://dx.doi.org/https://doi.org/10.1016/j.nima.2018.10.074}
  {\path{doi:https://doi.org/10.1016/j.nima.2018.10.074}}.
\newline\urlprefix\url{https://www.sciencedirect.com/science/article/pii/S0168900218313883}

\bibitem{Tsuji20-JINST}
N.~Tsuji, W.~Ootani, L.~Liu, K.~Yoshioka, Y.~Morita, M.~Gonokami,
  \href{https://doi.org/10.1088/1748-0221/15/05/c05052}{Study on saturation of
  {SiPM} for scintillator calorimeter using {UV} laser}, Journal of
  Instrumentation 15~(05) (2020) C05052--C05052.
\newblock \href {http://dx.doi.org/10.1088/1748-0221/15/05/c05052}
  {\path{doi:10.1088/1748-0221/15/05/c05052}}.
\newline\urlprefix\url{https://doi.org/10.1088/1748-0221/15/05/c05052}

\bibitem{vanDam10-IEEE}
H.~T. van Dam, S.~Seifert, R.~Vinke, P.~Dendooven, H.~Löhner, F.~J. Beekman,
  D.~R. Schaart, A comprehensive model of the response of silicon
  photomultipliers, IEEE Transactions on Nuclear Science 57~(4) (2010)
  2254--2266.
\newblock \href {http://dx.doi.org/10.1109/TNS.2010.2053048}
  {\path{doi:10.1109/TNS.2010.2053048}}.

\bibitem{Engelmann18-EPJC}
E.~Engelmann, E.~Popova, S.~Vinogradov, {Spatially Resolved Dark Count Rate of
  SiPMs}, Eur. Phys. J. C 78~(11) (2018) 971.
\newblock \href {http://dx.doi.org/10.1140/epjc/s10052-018-6454-0}
  {\path{doi:10.1140/epjc/s10052-018-6454-0}}.

\bibitem{Engelmann18-PhDTh}
E.~Engelmann, Dark count rate of silicon photomultipliers: Metrological
  characterization and suppression, Ph.D. thesis, Universität Hamburg (2018).

\bibitem{Otte17-NIMA}
A.~N. Otte, D.~Garcia, T.~Nguyen, D.~Purushotham,
  \href{https://www.sciencedirect.com/science/article/pii/S0168900216309901}{Characterization
  of three high efficiency and blue sensitive silicon photomultipliers},
  Nucl.~Inst.~Meth.~A 846 (2017) 106--125.
\newblock \href {http://dx.doi.org/https://doi.org/10.1016/j.nima.2016.09.053}
  {\path{doi:https://doi.org/10.1016/j.nima.2016.09.053}}.
\newline\urlprefix\url{https://www.sciencedirect.com/science/article/pii/S0168900216309901}

\bibitem{NOVO18-ARX}
K.~S. Ytre-Hauge, K.~Skjerdal, J.~Mattingly, I.~Meric, The novo project neutron
  detection for real-time range verification in proton therapy, a monte carlo
  feasibility study (2018).
\newblock \href {http://arxiv.org/abs/1810.07062} {\path{arXiv:1810.07062}}.

\end{thebibliography}

\end{document}